\documentclass[11pt]{article}
\usepackage[T1]{fontenc} 
\usepackage{color}
\usepackage{float}
\usepackage{amsmath}
\usepackage{amssymb}
\usepackage{setspace}
\usepackage{dsfont}
\usepackage[sort, authoryear]{natbib}

\usepackage{chngcntr}
\usepackage{caption}

\makeatletter

\usepackage{url}
\usepackage{booktabs}
\usepackage{array}
\usepackage{multirow}
\usepackage{float}

\usepackage{graphicx}
\usepackage{amsthm}
\usepackage{mathrsfs}

\theoremstyle{definition}
\newtheorem{result}{Result}[section]

\usepackage[toc,page]{appendix}

\usepackage{enumitem}
\setlist[enumerate]{topsep=0pt,itemsep=-1ex,partopsep=1ex,parsep=1ex}

\usepackage{parskip}
\usepackage{verbatim}
\usepackage{xcolor}
\usepackage{subcaption}

\definecolor{darkblue}{rgb}{0.0,0.0,0.55}
\RequirePackage[colorlinks,citecolor=darkblue,urlcolor=darkblue,linkcolor=darkblue]{hyperref} 

\usepackage[format=plain, labelfont=bf, singlelinecheck=off]{caption} 

\numberwithin{equation}{section}

\usepackage{algorithm}
\usepackage{algpseudocode}
\newlength{\continueindent}
\usepackage{etoolbox}

\newcommand*{\ALG@customparshape}{\parshape 2 \leftmargin \linewidth \dimexpr\ALG@tlm+\continueindent\relax \dimexpr\linewidth+\leftmargin-\ALG@tlm-\continueindent\relax}
\apptocmd{\ALG@beginblock}{\ALG@customparshape}{}{\errmessage{failed to patch}}

\algnewcommand{\algorithmicgoto}{\textbf{go to}}%
\algnewcommand{\Goto}[1]{\algorithmicgoto~\ref{#1}}%

\algdef{SE}[SUBALG]{Indent}{EndIndent}{}{\algorithmicend\ }%
\algtext*{Indent}
\algtext*{EndIndent}

\numberwithin{equation}{section}

\usepackage{xcolor}

\usepackage{geometry}

\allowdisplaybreaks

\newcommand{\bfa}{\mathbf{a}}
\newcommand{\bfb}{\mathbf{b}}

\newcommand{\bfh}{\mathbf{h}}

\newcommand{\bfk}{\mathbf{k}}

\newcommand{\bfr}{\mathbf{r}}
\newcommand{\bfs}{\mathbf{s}}

\newcommand{\bfw}{\mathbf{w}}
\newcommand{\bfx}{\mathbf{x}}
\newcommand{\bfy}{\mathbf{y}}
\newcommand{\bfz}{\mathbf{z}}

\newcommand{\bfA}{\mathbf{A}}
\newcommand{\bfB}{\mathbf{B}}

\newcommand{\bfD}{\mathbf{D}}

\newcommand{\bfF}{\mathbf{F}}

\newcommand{\bfH}{\mathbf{H}}
\newcommand{\bfI}{\mathbf{I}}

\newcommand{\bfM}{\mathbf{M}}

\newcommand{\bfR}{\mathbf{R}}
\newcommand{\bfS}{\mathbf{S}}

\newcommand{\bfU}{\mathbf{U}}
\newcommand{\bfV}{\mathbf{V}}
\newcommand{\bfW}{\mathbf{W}}

\newcommand{\bfY}{\mathbf{Y}}
\newcommand{\bfZ}{\mathbf{Z}}

\newcommand{\Ell}{\boldsymbol \ell}

\newcommand{\bfbeta}{\boldsymbol \beta}
\newcommand{\bfepsilon}{\boldsymbol \epsilon}
\newcommand{\bfdelta}{\boldsymbol \delta}

\newcommand{\bftheta}{\boldsymbol \theta}
\newcommand{\bfmu}{\boldsymbol \mu}

\newcommand{\bfgamma}{\boldsymbol \gamma}

\newcommand{\bfsigma}{\boldsymbol \sigma}

\newcommand{\bfSigma}{\boldsymbol \Sigma}

\newcommand{\bfOmega}{\boldsymbol \Omega}

\newcommand{\bfzero}{\mathbf{0}}

\newcommand{\GP}{\mathcal{GP}}
\newcommand{\MN}{\mathcal{MN}}
\newcommand{\MVN}{\mathcal{N}}

\newcommand{\Cjset}{\mathcal{C}_{t,j}}
\newcommand{\Cjcard}{|\mathcal{C}_{t,j}|}
\newcommand{\yObs}{\bfy^{\mathcal{D}}}
\newcommand{\wObs}{\bfw^{\mathcal{D}}}

\newcommand{\Expec}{\mathbb{E}}

\expandafter\def\expandafter\normalsize\expandafter{%
    \setlength\abovedisplayskip{1pt}
    \setlength\belowdisplayskip{6pt}
    \setlength\abovedisplayshortskip{1pt}
    \setlength\belowdisplayshortskip{6pt}
}

\makeatother

\theoremstyle{remark}

\newtheorem{proposition}{\textsc{\textbf{Proposition}}}[section]

\newcommand{\calX}{{\cal X}}

\defcitealias{NOAA}{NOAA2019}
\usepackage{setspace}

\newcommand{\blind}{1}

\begin{document}


\def\spacingset#1{\renewcommand{\baselinestretch}{#1}\small\normalsize} \spacingset{1}



\if1\blind
{
  \title{\bf Multivariate Gaussian process emulation for multifidelity computer models with high-dimensional spatial outputs}
  \author{Cyrus S. McCrimmon and Pulong Ma
   \hspace{.2cm}\\
    Department of Statistics, Iowa State University\\
    2438 Osborn Dr, Ames, IA 50011\\
} 
   \date{}
  \maketitle
} \fi

\if0\blind
{
  \bigskip
  \bigskip
  \bigskip
  \begin{center}
    {\LARGE\bf Multivariate Gaussian process emulation for multifidelity computer models with high-dimensional spatial outputs}
\end{center}
  \medskip
} \fi

\begin{abstract}
    Risk assessment of hurricane-driven storm surge relies on deterministic computer models that produce outputs over a large spatial domain. The surge models can often be run at a range of fidelity levels, with greater precision yielding more accurate simulations. Improved accuracy comes with a significant increase in computational expense, necessitating the development of an emulator which leverages information from the more plentiful low-fidelity outputs to provide fast and accurate predictions of high-fidelity simulations. To properly assess the risk of storm surge over a geographic region at aggregated spatial resolution, an emulator must account for spatial dependence between outputs yet remain computationally feasible for high-dimensional simulations. To address this challenge, we exploit the autoregressive cokriging framework to develop two cross-covariance structures to account for spatial dependence. One approach uses a separable covariance structure with a sparse Cholesky prior for the inverse of the cross-covariance matrix; the other involves a low-rank approximation via basis representations. We demonstrate their predictive performance in the storm surge application and a testbed example.
\end{abstract}

\noindent%

\newpage
\spacingset{1.58} 
\onehalfspacing 

\section{Introduction}

In science and engineering, computer models (also known as simulators) are used to simulate the behavior of physical systems. Computer models provide a digitized mathematical representation of a particular natural process. Most often, these computer models are very expensive to run to perform uncertainty quantification (UQ). There is extensive research in UQ that focuses on constructing surrogate models or emulators - fast probabilistic approximations to computer models \citep{santner2003design, gramacy2020surrogates}.

For complex physical systems, simulators are often run at a range of fidelity levels from low accuracy to high accuracy. Simulators at low-fidelity level (or low accuracy) are faster to run but less accurate than simulators at high-fidelity level (or high accuracy) because of differences in the complexity of the mathematical models, sophistication of the numerical solvers used and the resolution of meshes \citep{Peherstorfer2018}. The motivating application in this paper is a hurricane driven storm surge simulator featuring two fidelity levels. The ADCIRC ocean circulation model \citep{Luettich2004,westerink2008basin} is the primary computer model employed to study the abnormal rise in sea levels caused by storms, appearing in a number of applications, including the Federal Emergency Management Agency (FEMA) flood hazard map updates \cite[e.g.,][]{FEMA2008, niedoroda2010analysis, Jensen2012, Hesser2013} and in support of United States Army Corps of Engineers (USACE) projects \citep[e.g.,][]{Wamsley2013,Cialone2017}. The ADCIRC simulator is more accurate when coupled with the SWAN model, which incorporates wave effects, but a single high-resolution simulation in Southwestern Florida takes around 2,000 core-hours on a high-performance supercomputer \citep{xsede}. Risk assessment of storm surge requires a large number of model runs, making uncertainty quantification prohibitively costly. As a result, coastal flood hazard studies \citep[e.g.,][]{FEMA2015, FEMA2017} often use low-fidelity models without wave effects, although previous research \citep[e.g.,][]{dietrich2010high,marsooli2018numerical,yin2016coupled} has indicated the importance of using more comprehensive models to estimate storm surge.

To address this  scientific challenge, the goal of this paper is to develop a multifidelity emulator that synthesizes information from both low-fidelity simulator (ADCIRC) and high-fidelity simulator (ADCIRC + SWAN). There is extensive research on emulating a single computer model using Gaussian processes since the original work by \cite{Sacks1989}; see \cite{santner2003design} and \cite{ gramacy2020surrogates} for some background. The most widely adopted approach to emulate multifidelity computer models was proposed by  \cite{Kennedy2000}, who developed the so-called autoregressive cokriging model using a Markov property for the outputs at successive fidelity levels. Several extensions to the autoregressive cokriging model have been developed for various contexts \citep{Qian2008,Gratiet2013,Ma2020OBayes,Konomi2021}. The Markov property is appropriate for computer models where the outputs exhibit a natural ordering by fidelity level, as is the case for the aforementioned work. Important exceptions include non-ranked outputs \citep{Ji2024} and non-additive structures between fidelity levels \citep{Heo2025}, where \citet{Ji2024} emulate multifidelity computer models that may not be ranked from low to high fidelity levels, and \cite{Heo2025} emulate the nonlinear relationship between outputs at different fidelity levels. For the storm surge application, the ADCIRC + SWAN simulator has been validated against historical storms with improved performance compared to the ADICRC simulator \citep{Dietrich2012,Dietrich2011,dietrich2010high,FEMA2017,Cialone2017}. Our numerical studies also suggest that the Markov property is appropriate for the storm surge computer models.   

The ADCIRC and ADCIRC + SWAN models generate high-dimensional outputs over a large spatial region, producing the peak surge elevation at more than 9,000 spatial locations in the Cape Coral region of Florida, which is part of the FEMA study region \citep{FEMA2017}. The high-dimensional nature of the problem complicates the development of an emulator. The first challenge is to extend the aforementioned univariate approaches to accommodate high-dimensional outputs while maintaining computational efficiency. While we only consider a single spatial region with around $9,000$ locations in the storm surge application, the full mesh of the ADCIRC computer model can include millions of meshes when detailed surge outputs are needed for coastal flood hazard studies \citep{FEMA2015, FEMA2017}. This high-dimensional challenge is also broadly applicable for a wide range of computer simulations in physical sciences and engineering. The second challenge is to properly assess uncertainty quantification (UQ) at predicted outputs with a cross-covariance structure. Real-world storm surges often exhibit spatial dependence, where the storm surge for one location is similar to nearby locations. While incorporating the cross-covariance structure in an emulator is not strictly necessary to calibrate UQ at individual locations \citep[see][]{Gu2016, Ma2022PPCokrig}, this does not imply that predicted outputs at an aggregated spatial resolution will have proper UQ. In fact, if the outputs exhibit positive correlation in a spatial region (which is the case for storm surge), ignoring such cross-covariance structure would lead to under coverage for outputs at aggregated spatial resolutions; on the contrary, if the outputs exhibit negative correlation in a spatial region, ignoring the cross-covariance structure would lead to over-coverage at aggregated spatial resolutions. In the storm surge application, decision making often needs to be performed by jointly considering the surge predictions. For instance, storm surge risk assessment over a spatial region could be used to inform preparation for future storms, including the construction of defense structures \citep{FEMA2017}.  Scientific decisions sometimes need to be made at coarser spatial locations in terms of FEMA flood zones even though the computer model can predict storm surges at fine-scale spatial locations. Thus, it is of great importance to build an emulator that can account for the output cross-covariance to properly quantify uncertainty around predictions made at coarser spatial resolutions.

In the context of multivariate or high-dimensional spatial outputs, several modeling strategies have been developed in the literature. The first strategy is to ignore the difference between the output domain and the physical input domain, and then adopt computationally efficient Gaussian process (GP) models \citep[e.g.,][]{Gramacy2015, Katzfuss2021} on an extended domain by augmenting input domain with the output domain. While these GP approximations have been very successful for modeling univariate outputs when the underlying covariance structure is parametrically specified, typically with a stationary correlation function, this strategy oversimplifies the intrinsic nature of computer model outputs, and often there is no need for interpolation over the output domain. The second strategy is to assume independent models at each spatial location. The independence assumption allows model fitting in parallel, but it requires estimating a large number of model parameters, which is undesirable for model assessment when there is a massive number of outputs. A variant of this strategy is to use the parallel partial emulator \citep{Gu2016}, which assumes that the outputs are conditionally independent with common correlation parameters. However, this assumption may not be desirable when the joint distribution of outputs over a spatial domain is of primary interest. The third strategy is to consider a separable covariance structure \citep{Conti2010}, where the output cross-covariance does not change over the input space. The non-informative prior for the cross-covariance matrix in \cite{Conti2010} is infeasible when the number of model outputs is larger than the number of model runs, as the resulting posterior is improper. A remedy is to regularize the output covariance matrix, as was done in \citet{Mak2018} via the graphical lasso \citep{Friedman2008}. The fourth strategy is to use a nonseparable model \citep{Higdon2008,Bayarri2007}. This strategy treats outputs as a linear combination of basis functions and random weights. This strategy is closely related to the idea of the linear model of coregionalization (LMC) (\citeauthor{Matheron1982}, \citeyear{Matheron1982}; \citeauthor{Wackernagel2003}, \citeyear{Wackernagel2003}, pp.~175-176). Several variants for the construction of basis functions in the nonseparable model have also been proposed \citep[e.g.,][]{Guillas2018, Ma2022OSUE}. Among the four strategies, the separable and nonseparable cross-covariance structures are the most appealing for the storm surge application, where the output cross-covariance is of interest. Past work \citep{Fricker2013} has indicated that neither the separable nor the nonseparable model is universally dominant, and their predictive performance can vary by applications. 

In this paper, we propose a unifying Bayesian multivariate Gaussian process modeling framework for emulating multifidelity simulators with high-dimensional outputs using an autoregressive structure \citep{Kennedy2000}. The modeling framework is implemented using both the separable and nonseparable approaches described above. The first proposed Bayesian emulator features a separable covariance structure between the input space and the output space, with a sparse prior on the inverse cross-covariance matrix. We refer to this separable autoregressive cokriging model as \textit{SEP ARCokrig} (or simply SEP) hereafter. The main idea in SEP ARCokrig is to express the joint distribution of outputs using a series of successive multivariate regression models and then enforce conditional independence to induce sparsity in the inverse cross-covariance (or precision) matrix \citep[e.g.,][]{LeeLeeCholPrior, Bickel2008}. A prior is then assigned to the elements of the modified Cholesky decomposition of the precision matrix \citep{LeeLeeCholPrior}, which appear as parameters in the sequence of multivariate regression models, enabling computationally efficient Bayesian estimation of the inverse cross-covariance matrix. Our experience has indicated that posterior inference via the graphical lasso prior \citep{Wang2012} can be very slow due to the posterior sampling of the full precision matrix, and hence the graphical lasso prior is undesirable even when there is a moderate (e.g., a few thousand) number of outputs for a single model run. In the SEP ARCokrig model, we assign sparse Cholesky priors to the inverse cross-covariance matrix at each fidelity level. The resulting model thus exhibits a separable structure between the input space and  the output space and an autoregressive structure between fidelity level and input space. 

The second proposed Bayesian emulator, which has a nonseparable covariance structure, is based on the basis representation framework in \citet{Higdon2008}. We refer to this nonseparable autoregressive cokriging model as \textit{NONSEP ARCokrig} (or simply NONSEP) hereafter. At each fidelity level, the outputs are modeled as a linear combination of basis vectors and random weights. The random weights, which are modeled using Gaussian processes, are linked between fidelity levels via an autoregressive cokriging model so that nonseparability between the input space and the output space is maintained not only within fidelity levels but also across fidelity levels. This modeling strategy uses a Markov property for the latent processes across fidelity levels to maintain the assumed ordering of outputs by accuracy. Alternatively, if one jointly models the random weights across different fidelity levels via a multivariate GP model with a separable covariance structure, one actually obtains a separable cross-covariance in the output space, which may be undesirable. A crucial advantage in the proposed autoregressive cokriging modeling framework is that both SEP and NONSEP models can allow separate posterior inference for parameters at each fidelity level. We also show that at each fidelity level the regression parameters and variance parameters can be integrated out explicitly to allow fast statistical inference. This further implies that both SEP and NONSEP models can be easily extended to multiple fidelity levels without incurring extra computational bottlenecks since separate model fitting can be done. Such extension could be broadly applicable in various UQ applications.    

The rest of the paper is organized as follows. Section \ref{sec: model formulation} presents the proposed SEP and NONSEP models and their posterior inferences. Section \ref{sec: illustrations} demonstrates the performance of these two models in a testbed example for modeling the spread of chemical concentration and compares their performance with several existing models. Section \ref{sec: storm_surge_analysis} performs the multifideity emulation for the storm surge outputs in the Cape Coral region and compare their performance with existing models. Section~\ref{sec: discussion} concludes the paper with a summary and additional modeling perspectives. The supplement contains technical derivations and additional numerical demonstrations.

\section{Statistical Methodology} \label{sec: model formulation}

\subsection{Notation Setup}
Consider the collection of $m$ deterministic multifidelity computer models, sorted in ascending order by the accuracy of their representation to the physical process of interest. Each simulator generates $N$ outputs for a single input $\bfx$ in an input space denoted by $\calX\subset \mathbb{R}^{d}$. In the storm surge application, there are $m=2$ simulators with their outputs collected in a spatial domain in Southwest Florida, but the presented methods are applicable to more general settings. The computer model at fidelity level $t$ is run at a collection of $n_t$ inputs, denoted by $\calX_t$. We assume that the input design is nested, meaning $\calX_1 \supset \calX_{2} \supset \cdots \supset \calX_m$, as higher fidelity level means greater computational costs for running the computer model. In the storm surge application, the computational demand for the low-fidelity model ADCIRC is  several orders of magnitude smaller than the high-fidelity model ADCIRC + SWAN. The nested input design assumption can be satisfied by selecting inputs using a conditional Latin hypercube sample \citep{Minsny_clhs}.  

To simplify notation, let $y_{t,j}(\bfx)$ denote the output over input $\bfx$ at spatial coordinate $j$ and fidelity level $t$.
Let $\bfy_{t,j} = (\bfy_{t,j}(\bfx_1), \ldots, \bfy_{t,j}(\bfx_{n_t}))^{\top}$ be a vector of observed computer model outputs at coordinate $j$ and fidelity level $t$. Let $\bfy_t(\bfx):=(y_{t,1}(\bfx), y_{t,2}(\bfx), \ldots, y_{t,N}(\bfx))^\top$ denote a vector of $N$ outputs over input $\bfx$ at fidelity level $t$.  Let $\bfY_t =[\bfy_{t,1}, \ldots, \bfy_{t,N}]$ denote the $n_t\times N$ matrix of outputs across all inputs and spatial locations at fidelity level $t$. Let $\yObs = \{\bfY_1, \ldots, \bfY_m\}$ be the collection of corresponding outputs across all fidelity levels. 

Let $\bfz(\bfx):=(z_1(\bfx), \ldots, z_N(\bfx))^\top \in\mathbb{R}^N$ be a random vector. We denote $\bfz(\cdot):=\{\bfz(\bfx) \in\mathbb{R}^N: \bfx\in \calX\} \sim \mathcal{GP}(\mu(\cdot), r(\cdot, \cdot)\Sigma)$ if $\bfz(\cdot)$ follows a matrix-variate Gaussian process in $\mathbb{R}^N$ over input space $\mathcal{X}$, where $\mu(\cdot)$ denotes the mean function, $r(\cdot, \cdot)$ denotes the input correlation function, and $\bfSigma$ denotes the $N\times N$ cross-covariance matrix.  $\bfz(\cdot) \sim \mathcal{GP}(\mu(\cdot), r(\cdot, \cdot)\Sigma)$ if and only if for any $n$ inputs $\{\bfx_1,\ldots, \bfx_n\} \subset \calX$ and any integer $n>0$, the $n\times N$ random matrix $\bfZ:=[\bfz(\bfx_1), \ldots, \bfz(\bfx_n)]^\top \sim \mathcal{MN}_{n,N}( \bfmu, \bfR, \bfSigma)$ with $\bfmu:=[\mu(\bfx_1), \ldots, \mu(\bfx_n)]^\top$ and $\bfR:=[r(\bfx_i, \bfx_j)]_{i,j=1,\ldots, n}$; that is, $\bfZ$ follows a matrix-variate normal distribution \citep{Gupta1999} with the $n\times N$ matrix of means $\bfmu$ and $Cov(\text{vec}(\bfZ))=\bfSigma \otimes \bfR$. 

\subsection{Separable Autoregressive Cokriging} \label{sec: multivariate model}

\subsubsection{Model Formulation}
Based on the autoregressive cokriging framework, we assume the following multivariate autoregressive model: 
\begin{equation} \label{eqn: SEP cokriging} 
\begin{gathered}  
\bfy_{t}(\bfx)=\gamma_{t-1}\bfy_{t-1}(\bfx)+\bfdelta_{t}(\bfx),\quad t=2,\ldots,m, \\
\bfdelta_{t}(\cdot) \sim\mathcal{GP}(\bfh_{t}^{\top}(\cdot)\bfbeta_{t},\,r(\cdot,\cdot; \bftheta_{t})\bfSigma_t)\\
 \bfy_{1}(\cdot)  \sim\mathcal{GP}(\bfh_{1}^{\top}(\cdot)\bfbeta_{1},\,r(\cdot,\cdot; \bftheta_{1})\bfSigma_1),
\end{gathered}
\end{equation}
where $\bfdelta_{t}(\bfx)$ accounts for unexplained variability in the level-$t$ simulator outputs after linking successive models via the scale discrepancy parameter $\gamma_{t-1}$, and $\bfdelta_t(\bfx)$ is independent of $\bfy_1(\bfx), \ldots, \bfy_{t-1}(\bfx)$.  The mean functions in $\bfy_1(\cdot)$, $\bfdelta_2(\cdot), \ldots,$ $\bfdelta_m(\cdot)$ are modeled with  a $q_t$ dimensional vector of prespecified basis functions, $\bfh_t(\cdot)$, and a $q_t \times N$ matrix of regression coefficients, $\bfbeta_t$. The matrix-variate Gaussian processes $\bfy_1(\cdot)$ and $\bfdelta_t(\cdot)$ feature separable covariance structures \citep{Conti2010} with an input correlation function $r(\cdot, \cdot; \bftheta_t)$ and $N \times N$ cross-covariance matrix $\bfSigma_t$ that captures dependence of outputs over the output space. The model~\eqref{eqn: SEP cokriging} is an extension of the univariate autoregressive cokriging model \citep{Kennedy2000} for modeling multivariate computer models. We call this model a separable autoregressive cokriging (SEP) model.

At each fidelity level, we assume that a product form of correlation functions \citep{Sacks1989} is used  $r(\bfx,\bfx'; \bftheta) = \prod_{i=1}^{d} r^0(x_{i}, x'_{i} | \theta_{i})$, where $\bftheta$ denotes all the correlation parameters and $r^0(\cdot, \cdot|\theta)$ is chosen as the isotropic Mat\'ern correlation function of the form
\begin{align*}
    r^0(u | \theta) &= \frac{2^{1-\nu}}{\Gamma(\nu)}\left(\frac{\sqrt{2\nu}u}{\theta}\right)^{\nu}\mathcal{K}_{\nu}\left(\frac{\sqrt{2\nu}u}{\theta}\right),
\end{align*}
where $u$ is the Euclidean distance, $\Gamma(\cdot)$ is the gamma function, $\theta$ is the range parameter, $\mathcal{K}_{\nu}$ is the modified Bessel function of the second kind, and $\nu$ is the smoothness parameter that controls the mean-square differentiability of the random process. A standard practice is to fix $\nu$ at several values such as $\nu=2.5$ to allow closed-form formula. In our application, $\nu=2.5$ gives satisfactory results.     

\begin{figure}[htbp] 
\begin{subfigure}{.5\textwidth}
  \centering
\makebox[\textwidth][c]{ \includegraphics[width=.9\linewidth, height=0.2\textheight]{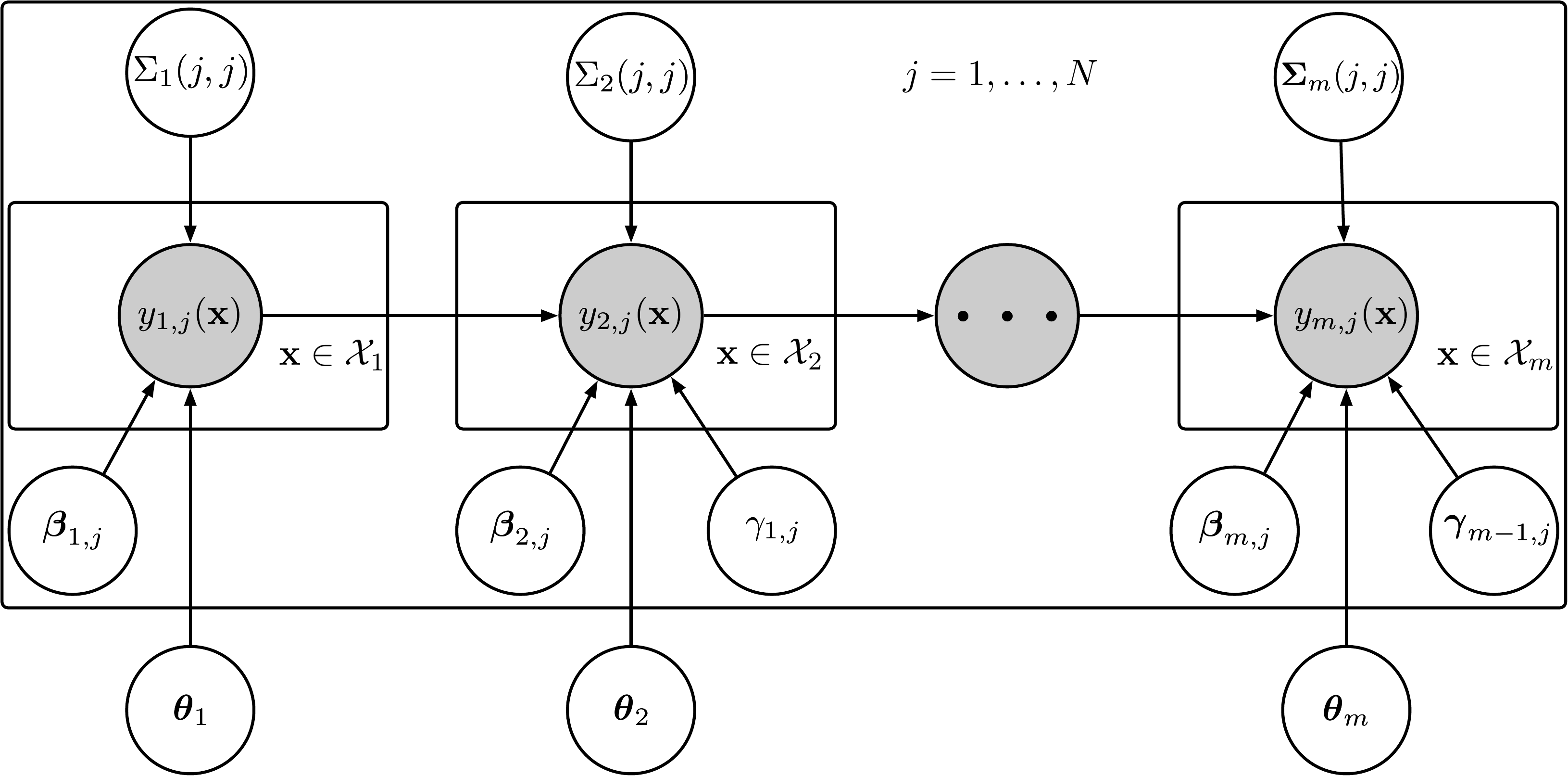}}
\caption{PP-ARCokrig}
\end{subfigure}%
\begin{subfigure}{.5\textwidth}
\makebox[\textwidth][c]{ \includegraphics[width=.9\linewidth, height=0.2\textheight]{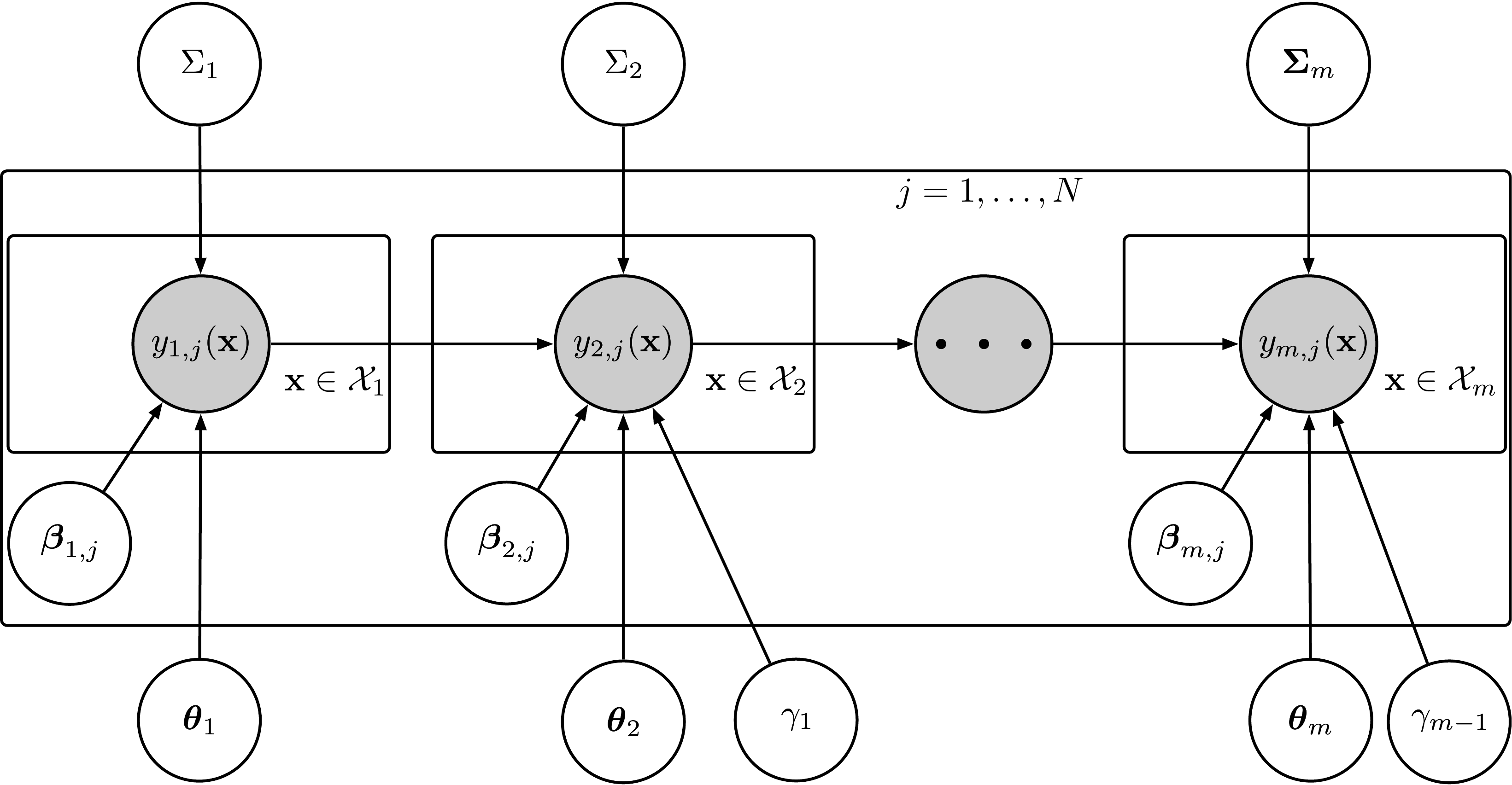}}
\caption{SEP}
\end{subfigure}
\caption{Graphical representation of two autoregressive cokriging models. Here $\Sigma_t(j,j)$ in the left panel is the $j$th diagonal element of the covariance matrix $\bfSigma_t$.}
\label{fig: graphical model for Sep-Cokriging}
\end{figure}

A special case of the separable autoregressive model is to assume a diagonal structure for $\bfSigma_t$, resulting in the so-called parallel partial (PP) autoregressive cokriging model \citep{Ma2022PPCokrig}. This is only appealing when the marginal predictive distribution at each spatial location is needed for uncertainty quantification and decision making. A graphical representation comparing the PP and SEP  models is given in Figure~\ref{fig: graphical model for Sep-Cokriging}. Capturing the dependence structure between outputs is critical when predictions  made at multiple outputs are jointly used for decision-making such as in the storm surge application. 

When $n_t\ll N$, estimating the dependence structure between outputs requires regularization in either the cross-covariance matrix $\bfSigma_t$ or the precision matrix $\bfOmega_t := \bfSigma_t^{-1}$. \cite{pourahmadi1999joint} proposed an autoregressive reparameterization of the precision matrix for a multivariate normal random vector that provides a natural method to construct a sparse representation of $\bfOmega_t$. We generalize this autoregressive reparametrization to a matrix-variate normal distribution. Our formulation is stated in the following result with its derivation given in Section~\ref{supp: MN_MCD} of the Supplementary Material. 

\begin{result} \label{res: MN_decom}
Let $\bfZ \sim \MN_{n,N}(\bfM, \bfR, \bfOmega^{-1})$ with mean $\bfM$, positive-definite correlation matrix $\bfR$ and cross-covariance matrix $\bfSigma$. Suppose that  the matrix $\bfOmega$ admits a modified Cholesky decomposition: $\bfOmega = (\bfI - \bfA)^{\top}\bfD^{-1}(\bfI-\bfA)$, where $\bfA$ and $\bfD$ are unique lower triangular and diagonal matrices, respectively. Let $\bfa_j$ be the vector of non-zero elements in row $j$ of $\bfA$ with $a_{jj}=0$ and $d_j>0$ be the $j$-th diagonal element of $\bfD$. Then 
    \begin{align}
    \begin{split}
        \bfZ_1 &\sim \MN_{n,1}(\bfM_1, \bfR, d_j)\\
        \bfZ_j | \bfZ_{1:j-1} &\sim \MN_{n,1}(\bfM_j + (\bfZ_{1:j-1} - \bfM_{1:j-1})\bfa_j, \bfR, d_j) \label{def: dist_y_j_MN}
        \end{split}
    \end{align}
where $\bfZ_j$ and $\bfM_j$ denote the $j$-th column of matrices $\bfZ$ and $\bfM$ and $\bfZ_{1:j-1}$ and $\bfM_{1:j-1}$ denote the submatrices of $\bfZ$ and $\bfM$ containing columns 1 to $j-1$.
\end{result}
\textbf{Result} \ref{res: MN_decom} shows that the problem of estimating the inverse cross-covariance matrix for a matrix-variate normal random variable $\bfZ$ is transformed into a series of $N$ autoregressive linear regression problems, where $\bfa_j$ and $d_j$ are the coefficients and variance, respectively, from regressing the centered elements of $\bfZ_j$ on the centered elements of $\bfZ_{1:j-1}$. If the columns of $\bfZ$ have a natural ordering, the inverse cross-covariance matrix can be approximated by regressing $\bfZ_j$ on the $p$ preceding columns to coordinate $j$ of $\bfZ$ \citep{Bickel2008, LeeLeeCholPrior}. This sparse representation of inverse cross-covariance matrix also implies conditional independence. Indeed, if $a_{ji} =0$ for $i<j$, then $\bfZ_{j}$ is conditionally independent of $\bfZ_i$ given $\bfZ_{1}, \ldots, \bfZ_{i-1}, \bfZ_{i+1}, \ldots, \bfZ_{j-1}$.   

For the separable autoregressive model~\eqref{eqn: SEP cokriging}, we assume that $\bfOmega_t:=\bfSigma_t^{-1}=(\mathbf{I}-\bfA_t)^\top \bfD_t^{-1}(\mathbf{I} - \bfA_t)$. Let $\bfa_{t,j}$ be the vector of non-zero elements in row $j$ of $\bfA_t$ and $d_{t,j}$ be the $j$-th diagnoal element of $\bfD_t$. Because the $N$ computer model outputs at each fidelity level were recorded at the same set of spatial locations, we can determine a sparse structure for $\bfA_t$ based on the proximity of one location to another. Let $\bfs_1, \ldots, \bfs_N$ be the spatial locations of the $N$ simulator outputs, sorted using the maximum-minimum-distance ordering \citep{Guinness_sharpening}. The vector of non-zero elements in row $j$ of $\bfA_{t}$ is defined as $\bfa_{t,j}^{(\mathcal{C}_{t, j})}$, and it includes all elements in the set $\{a_{t, ji}: i \in \mathcal{C}_{t,j}\}$, where $\mathcal{C}_{t, j}$ is a set comprised of the indices for the $p$ nearest spatial locations to $\bfs_{j}$ chosen from $\bfs_{1}, \ldots, \bfs_{j-1}$. When $p>j-1$, $\Cjset =\{1, \ldots, j-1\}$. In the storm surge application, the simulator outputs are very smooth over the spatial locations. We found that using a single nearest neighbor for each set $\Cjset$ works well. For applications where there is more variability between neighboring outputs, some regularization techniques \citep[see][]{kidd_katzfuss_BA2022} could be used to allow for larger neighbor sets in $\bfa_{t,j}^{\Cjset}$.  After constructing the neighbor sets, $y_{1,j}(\bfx)$ will be conditionally independent of $y_{1,i}(\bfx)$ for $i<j$ given $y_{1,1}(\bfx), \ldots, y_{1,i-1}(\bfx), y_{t, i+1}(\bfx), \ldots, y_{t, j-1}(\bfx)$ and the parameters in model~\eqref{eqn: SEP cokriging}. The same conditional independence assumption holds for levels $t>1$ given the outputs from level $t-1$.

 The collections of parameters in model~\eqref{eqn: SEP cokriging} across the $m$ fidelity levels are denoted by $\bfbeta:=\{\bfbeta_1, \ldots, \bfbeta_m\}$, $\bfgamma:=(\gamma_{1}, \ldots, \gamma_{m-1})^\top$, 
$\bftheta:=\{\bftheta_1, \ldots, \bftheta_m\}$, $\bfA := \{\bfA_1, \ldots, \bfA_m\}$ and $\bfD:=\{\bfD_1, \ldots, \bfD_m\}$. Because the input design is nested, it is easy to check that the marginal likelihood of the SEP model~\eqref{eqn: SEP cokriging} allows a closed-form expression
\begin{equation*} 
L({\bfy}^{\mathscr{D}} \mid \bfbeta, \bfgamma, \bftheta, \bfA, \bfD) 
=  \prod_{t=1}^m \mathcal{MN}_{n_t,N} (\bfY_t\mid \bfF_t \bfB_t, \bfR_t, \bfOmega_{t}^{-1}),
\end{equation*}
where for $t=1$, $\bfB_1: = \bfbeta_{1}$, $\bfF_1 = \bfH_1$, where $\bfH_t =\bfh_t(\calX_t)$ is an $n_t \times q_t$ matrix of the basis functions evaluated at $n_t$ inputs over $\mathcal{X}_t$. For $t>1$, $\bfB_t:=[\bfbeta_{t}^\top, \gamma_{t-1} \mathbf{I}_{N\times N}]^\top$, $\bfF_t = [\bfH_t, \bfU_{t-1}]$, where $\bfU_{t-1}:=[\bfy_{t-1,1}(\mathcal{X}_t),$ $\ldots, \bfy_{t-1,N}(\mathcal{X}_t)]$ is an $n_{t}\times N$ matrix of outputs from the model at level $t-1$ at $n_t$ inputs over $\mathcal{X}_{t}$. The level-$t$ correlation matrix  $\bfR_t:=r(\calX_t, \calX_t; \bftheta_t)$ has the $(i,j)$ element $r(\bfx_i, \bfx_j ; \bftheta_t)$ for $\bfx_i, \bfx_j \in \mathcal{X}_t$. 

For $j=1, \dots, N$ and $t=1, \ldots, m$, $\bfa_{t,j}^{(\Cjset)}$ and $d_{t,j}$ are assigned the following priors
\begin{align}
     \bfa_{t,j}^{(\mathcal{C}_{t,j})} | d_{t,j} \overset{\text{ind}}{\sim}\MVN_{\Cjcard}(\bfzero, d_{t,j}\tau^{2}_{t}\bfI), \quad
    d_{t,j} \overset{\text{ind}}{\sim} \text{IG}(\frac{\eta_{t}}{2}, \frac{\lambda_{t}}{2}), \label{def: prior ad}
\end{align}
where $\Cjcard$ is the cardinality of the set $\Cjset$, $\MVN(\cdot, \cdot)$ denotes a multivariate-normal distribution and $\text{IG}(\cdot, \cdot)$ denotes an Inverse-Gamma distribution. We assume that the hyperparameters $\tau^{2}_t$, $\eta_t$ and $\lambda_t$ are fixed. Similar versions of the prior in \eqref{def: prior ad} have been used to estimate the precision matrix of multivariate normal random variables in high-dimensional settings \citep{LeeLeeCholPrior, kidd_katzfuss_BA2022}. The joint prior distribution of $\bfbeta$, $\bfgamma$, $\bfA$, $\bfD$ and $\bftheta$ is  
\begin{align} \label{def: prior}
    \pi(\bfbeta, \bfgamma, \bfA, \bfD, \bftheta) & \propto \pi(\bfbeta, \bfgamma)\prod_{t=1}^{m}\{\pi(\bfA_{t}|\bfD_{t})\pi(\bfD_{t})\prod_{i=1}^d\pi(\theta_{t,i})\},
\end{align}
where $\pi(\bfbeta, \bfgamma) \propto 1$, $\pi(\bfA_{t} | \bfD_{t})$ and $\pi(\bfD_{t})$ are the priors defined in \eqref{def: prior ad}, while $\theta_{t,i} \overset{\text{iid}}{\sim}\mathcal{C}^{+}(0, q_{t,i})$, the half-Cauchy distribution with scale parameter $q_{t,i}$, for $i=1, \ldots, d$ and $t=1, \ldots, m$. In the numerical studies, $q_{t,i}$ is fixed at half the range of input domain along the $i$-th axis. The scale-discrepancy parameter $\gamma_{t-1}$ is often fixed at one due to identifiably issues. We also notice that allowing estimation of $\gamma$ does not improve the predictive accuracy in our numerical studies. In the following sections, we assume that $\bfgamma$ is fixed.

\subsubsection{Posterior Inference} \label{sec: posterior inference}

The form of the joint prior distribution in \eqref{def: prior} enables analytic integration of $\bfbeta$, $\bfA$ and $\bfD$ in $p(\bfbeta, \bfA, \bfD, \bftheta | \yObs)$ using standard conjugacy results, simplifying computation of the posterior distribution for computer models with high dimensional outputs. Let $\hat{\bfB}_1: = \hat{\bfbeta}_{t}$ and $\hat{\bfB}_t:=[\hat{\bfbeta}_{t}^\top, \gamma_{t-1} \mathbf{I}_{N\times N}]^\top$, where $\hat{\bfbeta}_{1} := (\bfH^{\top}_{1}\bfR^{-1}_{1}\bfH_{1})^{-1}\bfH^{\top}_{1}\bfR^{-1}_{1} \bfY_{1}$ and $\hat{\bfbeta}_{t} := (\bfH^{\top}_{t}\bfR^{-1}_{t}\bfH_{t})^{-1}\bfH^{\top}_{t}\bfR^{-1}_{t}(\bfY_{t} - \gamma_{t-1}\bfU_{t-1})$ for $t>1$ are the generalized least-squares estimates of $\bfbeta_{1}$ and $\bfbeta_{t}$. We state this result in the following proposition with its proof provided in Section \ref{supp: deriv_post_details} of the Supplementary Material. 
\begin{proposition}
The posterior distribution of $\bftheta$ is \label{prop: post_theta}
\begin{align} \label{def: post_theta}
    p(\bftheta | \yObs) \propto \prod_{t=1}^{m}|\bfR_{t}|^{\frac{-N}{2}} |\bfH^{\top}_{t}\bfR_{t}^{-1}\bfH_{t}|^{\frac{-N}{2}}(\hat{d}_{t,1})^{\frac{-\nu_t}{2}}\prod_{j=2}^{N} |\hat{\bfV}_{t,j}|^{\frac{1}{2}}(\hat{d}_{t,j})^{\frac{-\nu_t}{2}} \pi(\bftheta_t),
\end{align}
where $\nu_t = n_{t} +\eta_{t} -q_{t}$, $\hat{\bfV}_{t,j} = (\bfS_{t,\Cjset}^{\top}\bfS_{t,\Cjset} + \tau_t^{-2}\bfI)^{-1}$, $\hat{d}_{t,1} =(\bfS_{t,1}^{\top}\bfS_{t,1} + \lambda_{t})$, $\hat{d}_{t,t} = (\bfS_{t,j}^{\top} (\bfI - \bfS_{t,\Cjset}\hat{\bfV}_{t,j}\bfS_{t,\Cjset}^{\top})\bfS_{t,j} + \lambda_{t})$, $\bfS_{t} = \bfR^{\frac{-1}{2}}_{t}(\bfY_{t} - \bfF_{t}\hat{\bfB}_{t})$, and $\bfS_{t,\Cjset}$ is the submatrix of $\bfS_t$ containing all columns in the set $\Cjset$. 
\end{proposition}
Proposition~\ref{prop: post_theta} shows that the posterior distributions of model parameters $\bftheta_t$ are conditionally independent across different fidelity levels, facilitating fast computation. 
Posterior samples of $\bftheta_t$ are drawn separately for $t=1,\ldots, m$ using standard Metropolis-Hastings algorithms. Sampling from $p(\bftheta | \yObs)$ remains computationally feasible for large $N$ as long as the number of neighbors in each set $\Cjset$ is small. 

The primary goal of multifidelity computer model emulation is to determine the predictive distribution of the outputs for the simulator at highest fidelity level, $m$. For any new input $\bfx_0$, let $\bfy_{t}(\bfx_0):=(y_{t,1}(\bfx_0), \ldots, y_{t,N}(\bfx_0))^\top$ be a vector of $N$ outputs at fidelity level $t\in \{1,\ldots, m\}$ and new input $\bfx_0$.  Let $\bfy(\bfx_0):=(\bfy_{1}(\bfx_0)^\top, \ldots, \bfy_{m}(\bfx_0)^\top)^\top$ denote the collection of outputs at $\bfx_0$ across $m$ fidelity levels. 
We can establish that the posterior predictive distribution $p(\bfy(\bfx_{0}) | \bftheta, \yObs)$ has a closed-form expression with its proof given in Section \ref{supp: post_pred_details} of the Supplementary Material. 

Define  
\begin{align}
    \bfM_{t}(\bfx_0) &= \begin{cases}
    \hat{\bfbeta}^{\top}_{1}\bfh_{1}(\bfx_{0})  + (\bfY_{1} - \bfF_{1}\hat{\bfB}_{1})^{\top}\bfR_{1}^{-1}\bfr_{1}(\bfx_{0})&\text{ if } t=1\\
        \hat{\bfbeta}^{\top}_{t}\bfh_{t}(\bfx_{0}) + \gamma_{t-1}\bfy_{t-1}(\bfx_{0}) + (\bfY_{t} - \bfF_{t}\hat{\bfB}_{t})^{\top}\bfR_{t}^{-1}\bfr_{t}(\bfx_{0})&\text{ if } t>1
    \end{cases}  \label{def: M_t_star_star} \\
    R_{t}(\bfx_0) &= 1 - \bfr^{\top}_{t}(\bfx_{0})\bfR_{t}^{-1}\bfr_{t}(\bfx_{0})\notag\\
    &+ (h_{t}(\bfx_{0})- \bfH^{\top}_{t}\bfR_{t}^{-1}\bfr_{t}(\bfx_{0}))^{\top}(\bfH_{t}^{\top}\bfR_{t}^{-1}\bfH_{t})^{-1}(h_{t}(\bfx_{0})- \bfH^{\top}_{t}\bfR_{t}^{-1}\bfr_{t}(\bfx_{0})) \label{def: R_t_star_star},
\end{align}
where $\bfr_t(\bfx_0):=r( \bfx_0, \mathcal{X}_t; \bftheta_t)$ is an $n_t \times 1$ vector of correlations between $\bfx_0$ and the inputs in $\mathcal{X}_t$. 
\begin{proposition} \label{prop: S_pred_dist}
The distribution of $\bfy(\bfx_0) | \bftheta, \yObs$ is 
\begin{align} \label{dist: pred_prod_t}
    p(\bfy(\bfx_{0}) | \bftheta, \yObs)
    =& \prod_{t=1}^{m} \biggl\{ t_{\nu_{t}}(y_{t,1}(\bfx_{0})\mid \mu_{t,1}(\bfx_0), \sigma^{2}_{t,1}(\bfx_0)) 
    \prod_{j=2}^{N} t_{\nu_{t}}(y_{t,j}(\bfx_{0}) \mid \mu_{t,j}(\bfx_0), \sigma^{2}_{t,j}(\bfx_0)) \biggr\},
\end{align}
where $\mu_{t,1}(\bfx_0) = \bfM_{t,1}(\bfx_0)$, $\sigma^{2}_{t,1}(\bfx_0) = \nu_t^{-1} R_{t}(\bfx_0) \hat{d}_{t,1}$, and for $j>1$, $\mu_{t,j}(\bfx_0) = \bfM_{t,j}(\bfx_0) + (\bfy_{t,\Cjset}(\bfx_0) - \bfM_{t,\Cjset}(\bfx_0))^{\top}\hat{\bfa}_{t,j}^{(\Cjset)}$,
$\sigma^{2}_{t,j}(\bfx_0) = \nu_t^{-1} R_{t}(\bfx_0)\hat{d}_{t,j}((\bfy_{t,\Cjset}(\bfx_0) - \bfM_{t, \Cjset})^{\top}\hat{\bfV}_{t,j}(\bfy_{t,\Cjset}(\bfx_0) - \bfM_{t, \Cjset}) + 1)$, $\hat{\bfa}_{t,j}^{(\Cjset)} = \hat{\bfV}_{t,j}\bfS_{t,\Cjset}^{\top}\bfS_{t,j}$, and $\nu_t, \hat{d}_{t,j}, \hat{V}_{t,j}$ and $\bfS_{t}$ are defined in Proposition \ref{prop: post_theta}. 
\end{proposition}
Proposition~\ref{prop: S_pred_dist} shows that posterior samples from $ p(\bfy(\bfx_{0}) | \bftheta, \yObs)$ can be generated from a sequence of $m$ distributions. More importantly, because of the conditional independence assumption for the outputs, at each fidelity level, the posterior predictive distribution can be computed efficiently without factorizing a large cross-covariance matrix. Samples of $\bfy_{m}(\bfx_0)$ given $\yObs$ are obtained using typical Markov chain Monte Carlo (MCMC) algorithms. In addition, plug-in estimates for prediction via $p(\bfy(\bfx_{0}) | \bftheta, \yObs)$ is used to facilitate fast computation compared to a fully Bayes procedure. For instance, prediction can be obtained by plugging in $\bftheta$ with its posterior mode estimate $\hat{\bftheta}_{\text{MAP}}$ from the MCMC algorithm.


\subsection{Nonseparable Autoregressive Cokriging} \label{sec: nonsep cokriging}

\subsubsection{Model Formulation}
A separable covariance structure assumes the same cross-covariance in the output space, which may be restrictive in some real-world applications \citep[e.g.,][]{Fricker2013}. A nonseparable dependence structure is more flexible, but allowing nonseparability is particularly challenging for simulators that produce high-dimensional outputs. Following the idea in \cite{Higdon2008}, we also develop a nonseparable autoregressive model for multifidelity computer models.   
Similar to \cite{Higdon2008}, we first perform principal component analysis (PCA) on the outputs. In practice, this step is often performed on standardized outputs. For notational simplicity, we also refer to $\bfY_{t}$  as an $n_{t} \times N$ matrix of standardized outputs at fidelity level $t$. Then we assume a basis representation for the level-$t$ outputs 
\begin{align} \label{def: basis_rep}
 \bfy_{t}(\bfx) &=   \sum_{\ell=1}^{p_t} \bfk_{t,\ell} w_{t,\ell}(\bfx) + \bfepsilon_t,
 \end{align}
where $\bfk_{t,\ell}$ is the $\ell$-th $N\times 1$ principal component (or basis vector) constructed from principal component analysis of $\bfY_{t}$, $w_{t, \ell}(\bfx)$ is the random weight associated with the $\ell$-th basis vector, $p_t$ is the dimension of the reduced space and $\bfepsilon_t$ is a random reconstruction error. The reconstruction error, which accounts for discrepancies between the $p_t$ dimensional basis representation of the outputs and the actual values of $\bfY_{t}$, is modeled independently from the random weights. The reconstruction error and the selection of $p_t$ are discussed at the end of the section. 

For computer models with a single fidelity level, the random weights are often modeled using independent Gaussian processes \citep{Higdon2008}. To preserve the rank of output accuracy, we assume an autoregressive cokriging model for the PC weights. Let $\bfw_{t-1}(\cdot)=[w_{t-1,1}(\cdot),\ldots, w_{t-1,p_t}(\cdot)]^\top$ denote a vector of $p_t$ random functions whose partial realizations are given by PC scores.  In particular, we consider independent autoregressive cokriging models for the random weights across different PC weights: 
\begin{align} \label{mod: dist_wts}
        w_{t, \ell}(\cdot) | \bfw_{t-1}(\cdot), \bfbeta_{t-1, \ell}, \sigma^{2}_{t, \ell}, \bftheta_{t, \ell} & \overset{\text{ind}}{\sim} \GP(\bfbeta_{t-1}^\top \bfw_{t-1}(\cdot), \sigma^{2}_{t, \ell}r(\cdot, \cdot, \bftheta_{t, \ell})), \quad \ell=1,\ldots, p_t \notag \\
     w_{1, \ell}(\bfx) | \sigma^{2}_{1, \ell}, \bftheta_{1, \ell} & \overset{\text{ind}}{\sim} \GP(0, \sigma^{2}_{1, \ell}r(\cdot, \cdot;\bftheta_{1, \ell})),
\end{align}
where $\bfbeta_{t-1}:=[\beta_{t-1,1},\ldots, \beta_{t-1,p_t}]^\top$ denotes a vector of scale discrepancy parameters, $\sigma^{2}_{t,\ell}$ is a variance parameter and $r(\cdot, \cdot; \bftheta_{t,\ell})$ is a correlation function with parameter $\bftheta_{t,\ell}$. For $t=1,\ldots, m$ and $\ell=1,\ldots, p_t$, we used the same product form for the correlation function as the SEP model in \eqref{eqn: SEP cokriging}. The mean of the level-$1$ weights is set to $0$ because the outputs were centered before computing the principal components. The term $\bfbeta_{t-1}^\top \bfw_{t-1}(\cdot)$ acts like as a regression mean function with predictors given by the PC weight vectors $\bfw_{t-1}(\bfx_1), \ldots, \bfw_{t-1}(\bfx_{n_{t-1}})$. In the numerical experiment and application, we found that the PC weights $w_{t,\ell}(\bfx)$ is often highly correlated with the PC weights $w_{t-1,\ell}(\bfx)$. This implies that not all weights $\{w_{t-1,\ell}(\cdot), \ell=1,\ldots, p_{t-1}\}$  contribute to the variation of $w_{t-1, \ell} (\cdot)$. This actually can be used to improve model identifiability by only including those highly correlated weights at low-fidelity levels to capture the variability in the weights at high-fidelity level since the autoregressive cokriging model is not identifiable without informative prior on parameters \citep{Kennedy2001}. Nevertheless, prediction can still be carried out legitimately.   

\subsubsection{Posterior Inference} 
Let $\bfw_{t,\ell} = [w_{t,\ell}(\bfx_1), \ldots, w_{t,\ell}(\bfx_{n_t})]^{\top}$. Given the PC weights, the likelihood of model \eqref{mod: dist_wts} can be shown to have closed-form expression: 
\begin{align} \label{def: w_like}
    L(\wObs | \bfbeta_{w}, \bfsigma^{2}_{w}, \bftheta_{w}) &= \left\{\prod_{\ell=1}^{p_1}\MVN_{n_1}(\bfw_{1, \ell}| \bfzero, \sigma^{2}_{1, \ell}\bfR_{1, \ell}) \right\}
    \prod_{t=2}^{m}\prod_{\ell=1}^{p_t}\MVN_{n_t}\left(\bfw_{t,\ell}|  \bfW_{t-1} \bfbeta_{t-1, \ell}, \sigma^{2}_{t,\ell}\bfR_{t,\ell})\right), 
\end{align}
where $\wObs := \{\bfw_{1, 1}, \ldots, \bfw_{m, 1}, \ldots, \bfw_{m, p_m}\}$, $\bfbeta_{w} = \{\bfbeta_{1, 1}, \ldots, \bfbeta_{m-1, p_{m-1}}\}$, $\bfsigma^{2}_w := \{\sigma^{2}_{1, 1},\ldots,  \sigma_{m,p_m}^2 \}$, $\bftheta_w := \{\bftheta_{1,1}, \ldots, \bftheta_{m, p_m}\}$, $\bfW_{t-1}$ is an $n_t \times p_{t-1}$ matrix of PC weights at the level $t-1$  over the inputs in $\mathcal{X}_t$ and $\bfR_{t, \ell}$ is an $n_t \times n_{t}$ correlation matrix with the $(i,j)$ element $r(\bfx_i, \bfx_j;\bftheta_{t, \ell})$. To complete the Bayesian specification, we assume the following prior distribution for  $\bfbeta_{w}, \bfsigma^{2}_w, \bftheta_{w}$ 
\begin{align} \label{def: w_prior}
    \pi(\bfbeta_{w}, \bfsigma^{2}_w, \bftheta_w) & \propto
    \prod_{t=1}^{m}\prod_{\ell=1}^{p_t}\frac{1}{\sigma^{2}_{t, \ell}}\pi(\bftheta_{t,\ell}), 
\end{align}
where $\pi(\bftheta_{t, \ell})$ is the density for the prior on correlation parameter $\bftheta_{t,\ell}$. We used independent half-Cauchy priors for the range parameters at $t=1$. For $t>1$, we found that more informative prior distributions for the range parameters help stabilize posterior sampling with independent half-normal prior distributions. Using standard conjugacy results, the likelihood~\eqref{def: w_like} and prior~\eqref{def: w_prior} enable analytic integration of $\bfbeta_w$ and $\sigma^{2}_w$ from the posterior $p(\bfbeta_w, \bfsigma^{2}_w, \bftheta_w | \wObs)$, yielding a closed-form expression for $p(\bftheta_w | \wObs)$. Samples from the posterior distribution $p(\bftheta_w | \wObs)$ are generated using MCMC (Markov chain Monte Carlo) algorithms. 

For new input $\bfx_0$, we also derived a closed-form expression for the predictive distribution $p (\bfw(\bfx_0) | \bftheta_w, \wObs)$, where $\bfw(\bfx_0):=(\bfw_{1}(\bfx_0)^{\top}, \ldots, \bfw_{m}(\bfx_0)^{\top})^{\top}$. The expressions for both $p(\bftheta_w | \wObs)$ and $p(\bfw(\bfx_0) | \bftheta_w, \wObs)$ are provided in Section \ref{supp: sup_nonsep} of the Supplementary Material. Similarly, we noticed that separate modeling fitting and prediction can be done because of the autoregressive structure. To predict the $m$-th centered and scaled computer model outputs for $\bfx_0$, posterior samples drawn from $p(\bfw_m(\bfx_0) | \wObs)$ are projected back to the original space using the basis representation in \eqref{def: basis_rep} with 
\[
\bfepsilon_m = \sum_{\ell=p_m+1}^{n_m}  \bfk_{m, \ell}\epsilon_{m,\ell},
\] 
where $\bfk_{m,\ell}$ is the $\ell^{\text{th}}$ discarded principal component direction and the $\epsilon_{m,\ell}$ are independent normal random variables with variance equal to the variance of $\bfY_m$ explained by direction $\bfk_{m,\ell}$ \citep{wilkinson_pca}. Predictions are then transformed to the original scale of $\bfY_m$. While the primary goal of multifidelity computer model emulation is to predict the outputs at the highest fidelity level, the basis representation~\eqref{def: basis_rep} enables predictions at any fidelity level. To select the number of basis vectors,  \cite{Higdon2008} found that choosing the first few principal components to explain greater than $95$ or $99$ percent of the total variation in the data works well. However, in some applications, the remaining principal components that explain a very small percentage of total variability could add meaningful contributions to predictions \citep{wilkinson_pca}. To guide the selection of basis vectors, we evaluate the root mean square prediction error (RMSPE) in the model based on different number of basis vectors in the numerical studies.

\section{TestBed Example} \label{sec: illustrations}

We study the predictive performance of the SEP and NONSEP emulators using a simulated computer experiment from \cite{Bliznyuk_env} As a comparison to the proposed methods, we also fit two models that feature independent Gaussian processes at each location with shared correlation parameters: the parallel partial autoregressive cokriging (PP-ARCokrig) model  \citep{Ma2022PPCokrig} via the \textsf{R} package \texttt{ARCokrig} \citep{ARCokrig} and the parallel partial (PP) model \citep{Gu2016} via the \textsf{R} package \texttt{RobustGaSP} \citep{Gu2018RobustGP}. The PP emulator is fit to the high-fidelity outputs only. We also fit the sparse Vecchia approximation \citep{Katzfuss2021} for the data over the product space of the input space and spatial domain, but the results are omitted due to worse predictive performance. To assess the predictive performance of each emulator at multiple outputs, we also consider the average high-fidelity output for input $\bfx$
\begin{align} \label{def: y_bar_2}
    \Bar{y}_{2}(\bfx) &= N^{-1}\sum_{j=1}^{N}y_{2,j}(\bfx),
\end{align}
which is useful when decision-making is needed at aggregated spatial resolutions. In what follows, we illustrate the predictive performances among these four models: PP, PP-ARCokrig, SEP, and NONSEP, where PP is fitted to the high-fidelity data only. To measure predictive performance, we use RMSPE and the coverage probability of $95\%$ equal-tailed credible intervals (CVG), and average length of $95\%$ equal-tailed credible intervals (ALCI).

\subsection{Experimental Design} 
In this testbed example, we consider the computer model from \cite{Bliznyuk_env} that captures the spread of a contaminant after a chemical spill over a space-time domain using the function
    $y_2(\bfx, \bfs) = \sqrt{4\pi}C(\bfx, \bfs),$
where $C(\bfx, \bfs)$, the concentration of the pollutant evaluated at the input $\bfx :=(M, D, L, T)^{\top}$ and space-time coordinate $\bfs:=(s_{1},s_{2})$, is
\begin{align} \label{def: C}
    C(\bfx, \bfs) &= \frac{M}{\sqrt{4\pi Ds_{2}}}\exp\left(\frac{-s_{1}^{2}}{4Ds_{2}}\right) + \frac{M}{\sqrt{4\pi D(s_{2}-T)}}\exp\left\{-\frac{(s_{1}-L)^{2}}{4D(s_{2}-T)}\right\} \mathds{1}(T < s_{2}),
\end{align}
where the parameter $M \in [7, 13]$ is the mass of the pollutant spilled at each location, $D\in [0.02, 0.12]$ is the diffusion rate in the channel, $L\in[0.01,3]$ is the location of the second spill and $T \in [30, 30.295]$ is the time of the second spill, $s_1$ is the spatial coordinate and $s_2$ is the time. 

To mimic a multifidelity computer model, the function $y_2(\bfx, \bfs)$ is approximated using a one-term Taylor series expansion of the two exponential terms in Equation~\eqref{def: C}. In what follows, we refer to $y_1(\bfx, \bfs)$, the approximation of $y_2(\bfx, \bfs)$, as the low-fidelity model. For each input setting, the low and high fidelity simulators were evaluated at a grid of 20 evenly spaced-spatial coordinates over $[0.5, 5]$ and 50 evenly-spaced time points over $[35,60]$, respectively. With the parameter $T$ fixed at $30$, $60$ values of $(M, D, L)$ were selected for the low-fidelity model using a Latin hypercube sample \citep{gramacy2020surrogates}. The function $y_2(\bfx, \bfs)$ was evaluated at $30$ inputs drawn from the collection of $60$ low-fidelity inputs using a conditional Latin hypercube sample \citep{Minsny_clhs}. To assess the predictive performance, $30$ test inputs of $(M,D,L)$ were randomly sampled using a uniform distribution over $[7,13] \times [0.02, 0.12] \times [0.01,3]$.

\subsection{Emulation Results} 

Before we fit the models, it is often helpful to perform diagnostic checks to identify key data analytic challenges and rule out implausible models. The empirical mean and variance of the outputs at each spatial location indicate that the output space is nonstationary; see Figure~\ref{fig: EDA testbed} of the Supplementary Material. Thus, a stationary output covariance structure is not appropriate for the application. Instead, we may wish to fit PP, PP-ARCokrig, SEP, and NONSEP. After taking the difference between the high-fidelity outputs and low-fidelity outputs, the residuals show much less variability than the original outputs but still exhibit a high degree of spatial correlation, suggesting that the autoregressive structure is reasonable for the application. 

\begin{table}[htb!]
\centering
\caption{Predictive performance of different models at $30$ test inputs and $1,000$ spatial locations. Predictive performance is evaluated for individual high-fidelity outputs and for the aggregated high-fidelity outputs.}
\label{tab: sim_pred}
{\resizebox{1.0\textwidth}{!}{%
\begin{tabular}{lcccccc}
\toprule
  & \multicolumn{3}{c}{Marginal predictive performance of $y_{2,j}(\cdot)$} & \multicolumn{3}{c}{Predictive performance of $\Bar{y}_2(\cdot)$} \\
\cmidrule(lr){2-4} \cmidrule(lr){5-7}
 Model & RMSPE & CVG ($95\%$) & ALCI ($95\%$)
 & RMSPE & CVG ($95\%$) & ALCI ($95\%$) \\
\midrule
SEP & 0.195      & 97.0               & 0.739 & 0.051      & 93.3              & 0.158         \\
NONSEP & 0.250      & 97.3               & 0.870 & 0.049       & 100              & 0.402         \\
PP-ARCokrig & 0.208      & 95.7               & 0.626 & 0.045      & 23.3               & 0.023        \\
PP & 0.267       & 95.9               & 0.726 & 0.058      & 20              & 0.028          \\ 
\bottomrule
\end{tabular}
}}
\end{table}

We show the predictive performance for the four aforementioned models in Table~\ref{tab: sim_pred}. The metrics for marginal predictive performance in Table~\ref{tab: sim_pred} are computed at individual spatial locations and then averaged over the entire spatial domain. In terms of RMSPE, the SEP model achieve the best results. In terms of CVG and ALCI, all models achieve good performance. While PP and PP-ARCokrig do not account for output cross-covariance structure, it is not surprising that they can achieve calibrated CVG and ALCI since these metrics evaluate the performance of marginal predictive distributions at individual spatial locations. When comparing the performance at aggregated spatial locations, such as the spatial average defined in Equation~\eqref{def: y_bar_2}, we expect the performance of PP and PP-ARCokrig to deteriorate substantially because the two models do not allow proper UQ, which is confirmed by the results in Table~\ref{tab: sim_pred}. Among the two proposed models, SEP gives better results than NONSEP in terms of UQ. The detailed implementation of SEP is given in Section~\ref{supp: SEP_testbed} of the Supplementary Material.

\begin{figure}[ht!]
    \centering
    \includegraphics[width=0.35\textwidth, height=0.25\textheight]{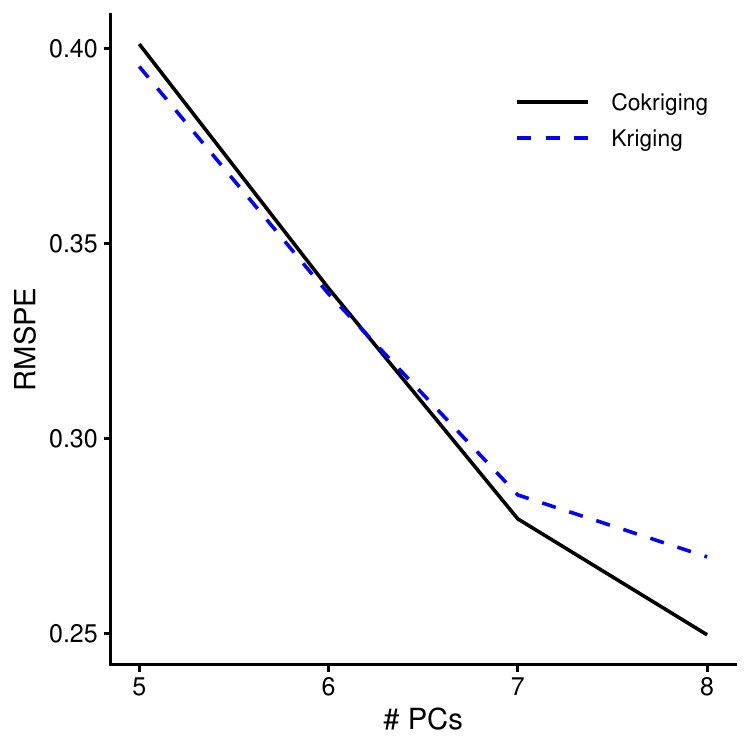}
    \caption{RMSPE as a function of number of principal components. The black line represents the results under the NONSEP cokriging model while the blue-dashed line represents the results based on NONSEP kriging model fit to the high-fidelity simulator only. The RMSPE was computed using the average posterior predictive weight.}
    \label{fig: RMSPE_sim_nonsep_lev}
\end{figure}

In the NONSEP model, we performed PCA separately for the outputs at each fidelity level. Figure~\ref{fig: sim_scatter_wts} of the Supplementary Material shows that PC scores at different fidelity levels are only highly correlated when they share the same component number, indicating that any PC weight function at the high-fidelity level can just be modeled with the corresponding weight function at the low-fidelity level. The reason for this is likely due to the good approximation of low-fidelity simulator to the high-fidelity simulator. The first four PCs explain more than $99\%$ of the total output variability at each fidelity level, but the NONSEP model does not give satisfactory results with only four PCs; in contrast, choosing the number of PCs based on RMSPE would lead to substantial improvement in prediction; see Figure~\ref{fig: RMSPE_sim_nonsep_lev}. Because the RMSPE plateaus after an eighth principal component is added to the model, we chose eight PCs for the basis representation~\eqref{def: basis_rep} at both levels to fit the autoregressive cokriging model~\eqref{mod: dist_wts}. For each PC weight and fidelity level, the Metropolis-Hastings algorithm was run for $60,000$ iterations (with a burn-in of $6,000$ samples) to obtain posterior samples of the correlation parameters. We noticed that the MCMC samples do not show any lack of convergence behaviors and posterior distributions are not sensitive to different hyperparameter settings. Figure~\ref{fig: RMSPE_sim_nonsep_lev} shows how the RMSPE for the NONSEP emulator (cokriging) compares to a model fit using only the PC weights from level-2 (kriging). The autoregressive structure for the PC weights improves predictions of the cokriging model relative to the kriging model for the seventh and eighth principal components, where there is a weaker relationship between the random weights and inputs.

\section{Multifidelity Emulation of Storm Surges} \label{sec: storm_surge_analysis}

\subsection{Storm Surge Simulators and Experimental Design} 
ADCIRC is an ocean circulation model used to simulate hurricane driven storm surge \citep{Luettich2004}. ADCIRC is more accurate when coupled with the SWAN model \citep{Booij1999, Zijlema2010}, which incorporates wave effects. This computer model was studied by \citep{Ma2022PPCokrig}, where  downstream applications and validation of computer models have been introduced.  We consider the peak surge elevation, measured in meters (m), generated at $9,284$ locations in the Cape Coral region that is selected based on the sea levels that are of primary interest in coastal flood hazard studies. The ADICRC and ADCIRC + SWAN simulators take six input parameters: $\Delta P \in [30,70]$ (central pressure deficit (mb)), $R_p \in [16,39]$ (scale pressure radius (nautical miles)), $V_f \in [3,10]$ (forward speed (m/s)), $\theta \in [15, 75]$ (heading in degrees clockwise from north), $B \in [0.9,1.4]$ (Holland's B parameter), $\mathbf{\Ell}$ (landfall location in latitude and longitude). The ranges of these parameters are selected based on the characteristics of landfalling hurricanes.  

Following \cite{Ma2022PPCokrig}, experimental design is based on a maximin Latin hypercube design to select $50$ unique combinations of $(\Delta P, R_p, V_f, \theta, B)$ that characterize storm characteristics. For each storm parameter of $(\Delta P, R_p, V_f, \theta, B)$, the landfall location $\ell$ was chosen with one-half of $R_p$ uniformly spaced over the coastline but with random initial landfall locations on the southwest position; see \cite{Ma2022PPCokrig} for additional details. In total, the ADCIRC and ADCIRC + SWAN computer models were run at 226 inputs after excluding unrealistic storm characteristics. Since the coastline in Cape Coral has very low variation in latitude, $\ell$ was converted to a distance measure, $d_\ell$, using the location farthest to the northwest along the coastline as a reference.  In terms of training, random samples of 120 and 60 inputs were selected for the ADCIRC and ADCIRC + SWAN simulators, respectively, using a nested design. The remaining 106 inputs are used to assess predictive performance at individual spatial locations and aggregated spatial resolutions. 

\subsection{Diagnostics} 
As in the testbed example, we first perform exploratory data analysis to check model assumptions. Figure~\ref{fig: ss_mean_var_res} of the Supplementary Material shows that there is region-specific spatial heterogeneity in mean and variance for both low and high fidelity outputs, indicating that models like PP and PP-ARCokrig might be suitable. The residuals in Figure~\ref{fig: ss_mean_var_res} indicates that after taking out the variability due to low-fidelity simulation, the high-fidelity simulation shows some region specific nonstationary behaviors, which favors a nonstationary covariance structure as in the four models under comparison: PP, PP-ARCokrig, SEP, and NONSEP. Because nearby outputs tend to be similar (see Figure~\ref{fig: ss_mean_var_res}), the output cross-covariance must be accounted for. As such, models like SEP and NONSEP, which include explicit cross-covariance structures are likely to improve model fitting. 

Figure~\ref{fig: ss_mean_var_res} of the Supplementary Material shows spatial maps of the empirical mean and empirical variance in the residuals (difference between high-fidelity simulator and low fidelity simulator) over nested inputs, indicating that sensible models need to account for spatially-varying patterns in the mean function and cross-covariance matrix in the output space.  The residual plots in Figure~\ref{fig: ss_mean_var_res} of the Supplementary Material also imply that the autoregressive structure is a reasonable assumption because the variation in the high-fidelity simulator is substantially reduced after regressing on the low-fidelity simulator for overlapping inputs. As shown in Figure~\ref{fig: storm surge PC weights} of the Supplementary Material, there is a very strong linear trend between low-fidelity PC scores and high-fidelity PC scores at the same PC component number, which allows us to specify the mean functions in the autoregressive structure in model~\eqref{mod: dist_wts}.

\subsection{Emulation Results} 
\begin{figure}[hbt!]
    \centering
    \begin{subfigure}[b]{0.49\textwidth}
        \centering
        \includegraphics[width=\textwidth,height=0.2\textheight]{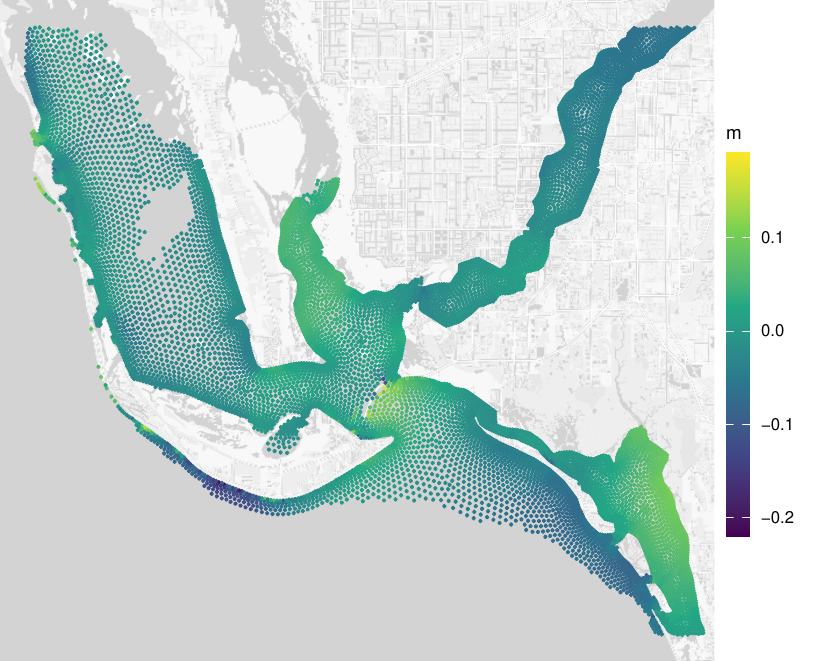}
        \caption{Difference SEP.}
    \end{subfigure}
    \hfill
    \begin{subfigure}[b]{0.49\textwidth}
        \centering
        \includegraphics[width=\textwidth,height=0.2\textheight]{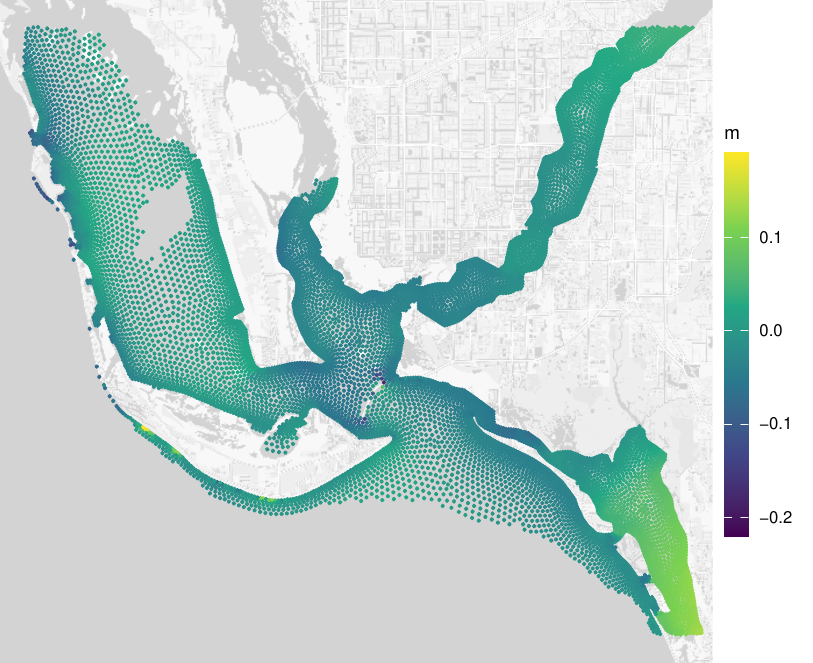}
        \caption{Difference NONSEP.}
    \end{subfigure}
    \vspace{0.5cm}
    \begin{subfigure}[b]{0.49\textwidth}
        \centering
        \includegraphics[width=\textwidth,height=0.2\textheight]{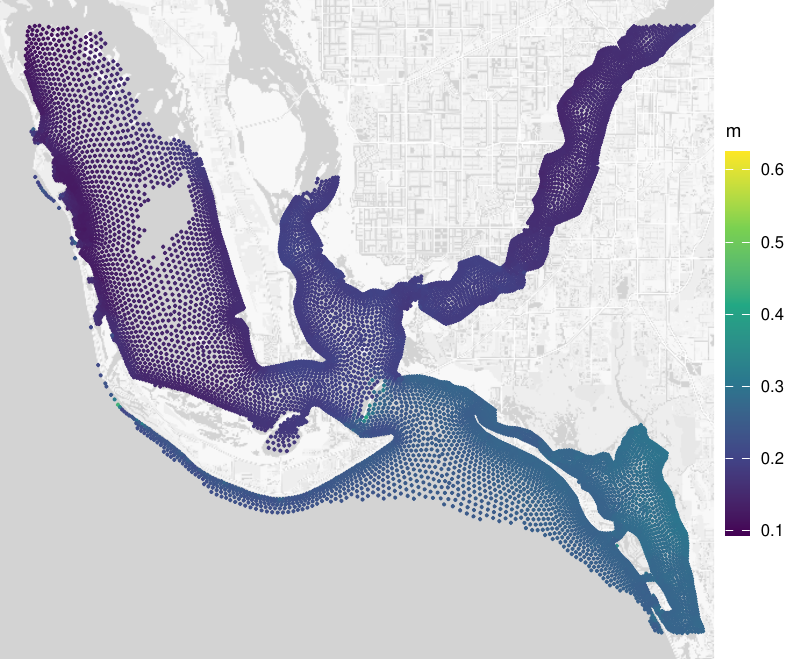}
        \caption{CI Len. SEP.}
    \end{subfigure}
    \hfill
    \begin{subfigure}[b]{0.49\textwidth}
        \centering
        \includegraphics[width=\textwidth,height=0.2\textheight]{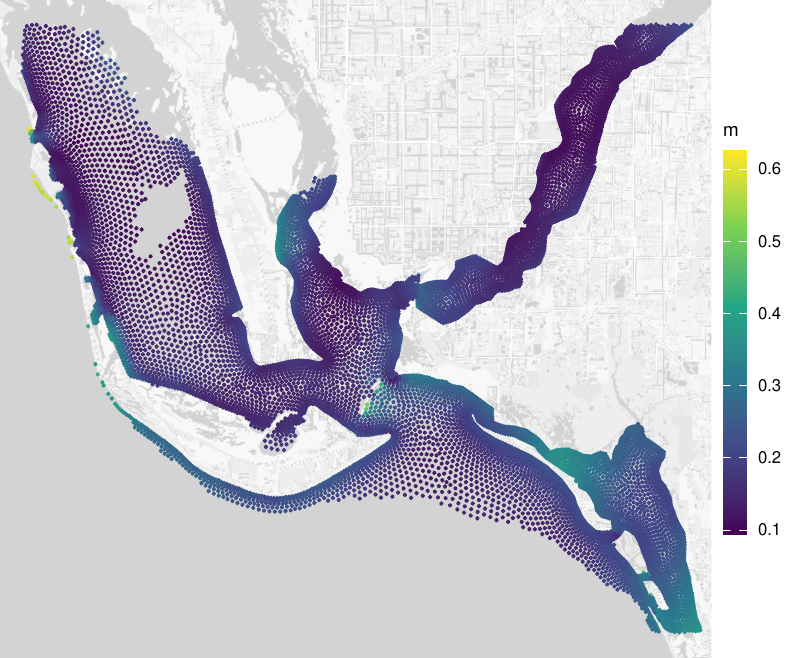}
        \caption{CI Len. NONSEP.}
    \end{subfigure}
    
    \caption{Predictive comparison under SEP and NONSEP over one testing input: $\Delta P = 37.58, R_p = 29.37, V_f =7.96, \theta=70.42, B=1.38, \ell =(-82.25, 26.83)$. Panel (a) shows the difference between actual high-fidelity output and predicted output under the SEP model; Panel (b) shows the difference between high-fidelity output and predicted output under the NONSEP model. Panels (c) and (d) show the lengths of 95\% equal-tail credible interval under the SEP and NONSEP models, respectively.}
    \label{fig: ss_pred_comp}
\end{figure}

The predictive performance over four different models are given in Table~\ref{tab: storm surge prediction}. In terms of RMSPE, PP-ARCokrig gives the best results for assessment over individual location and aggregated spatial locations. This is not surprising since past work \citep{Fricker2013} found that including an output dependence structure in a multivariate emulator may lead to slightly worse marginal predictive performance  in terms of RMSE. However, we can see from Table~\ref{tab: storm surge prediction} that both SEP and NONSEP based on seven PCs give better UQ in terms of CVG ($95\%$) and ALCI ($95\%$). This agrees with the findings in \cite{Mak2018}. When UQ is assessed for predictive distributions over aggregated quantities, the lack of cross-covariance in models such as PP and PP-ARCokrig becomes a serious issue as indicated in Table~\ref{tab: storm surge prediction}. Compared with SEP, NONSEP gives better UQ of predictive distributions at individual locations and aggregated spatial resolution. Figure~\ref{fig: ss_pred_comp} also confirms that the NONSEP model can better capture the residuals and prediction uncertainty is smaller in smooth spatial regions. The detailed implementation of the SEP model is given in Section~\ref{supp: storm surge} of the Supplementary Material. In what follows, we focus on presenting the results on the NONSEP model because of its superior performance.

\begin{table}[hbt!] 
\centering
\caption{Predictive performance of ADCIRC + SWAN emulators at $106$ test inputs and $9,284$ spatial locations. }
\label{tab: storm surge prediction}
{\resizebox{1.0\textwidth}{!}{%
\begin{tabular}{lcccccc}
\toprule
  & \multicolumn{3}{c}{Marginal predictive performance of $y_{2,j}(\cdot)$} & \multicolumn{3}{c}{Predictive performance of $\Bar{y}_2(\cdot)$} \\
\cmidrule(lr){2-4} \cmidrule(lr){5-7}
 Model & RMSPE & CVG ($95\%$) & ALCI ($95\%$)
 & RMSPE & CVG ($95\%$) & ALCI ($95\%$) \\
\midrule
SEP & 0.111      & 91.50               & 0.409 & 0.079     & 90.57               & 0.284        \\
NONSEP & 0.102      & 94.65               & 0.373 & 0.060      & 96.23              & 0.248       \\     

PP-ARCokrig & 0.099      & 91.32               & 0.345 & 0.057      & 4.72               & 0.004        \\

PP & 0.170      & 84.07               & 0.452 & 0.118      & 0.94               & 0.005        \\

\bottomrule
\end{tabular}
}}
\end{table}

\begin{figure}[hbt!]
    \centering
    \begin{subfigure}[b]{0.32\textwidth}
        \centering
        \includegraphics[width=\textwidth]{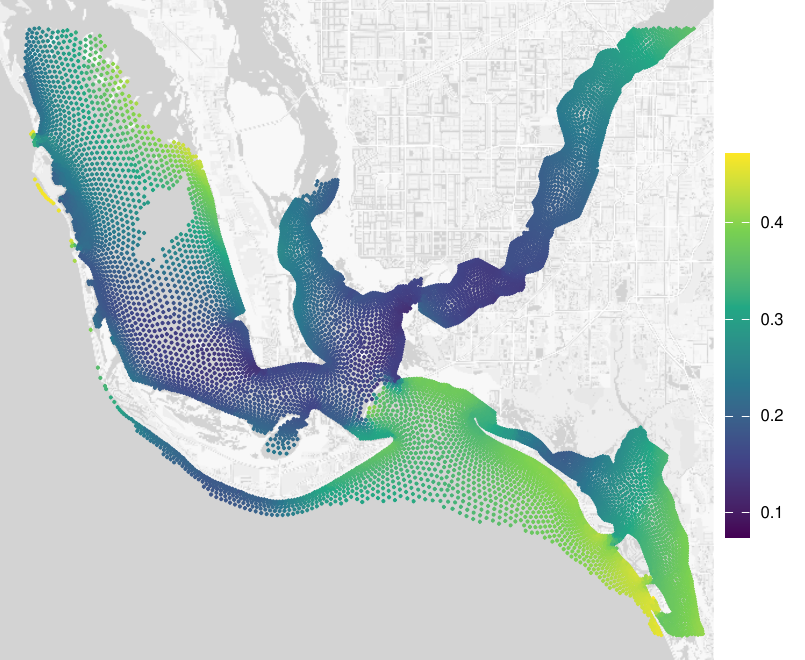}
        \caption{RMSPE (1 PC)}
    \end{subfigure}
    \hfill
    \begin{subfigure}[b]{0.32\textwidth}
        \centering
        \includegraphics[width=\textwidth]{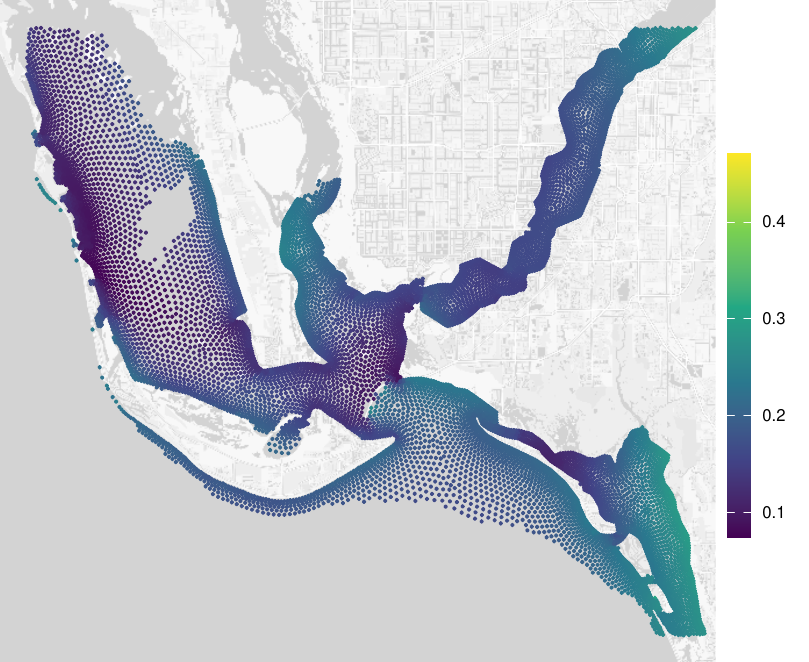}
        \caption{RMSPE (2 PCs)}
    \end{subfigure}
    \hfill
    \begin{subfigure}[b]{0.32\textwidth}
        \centering
        \includegraphics[width=\textwidth]{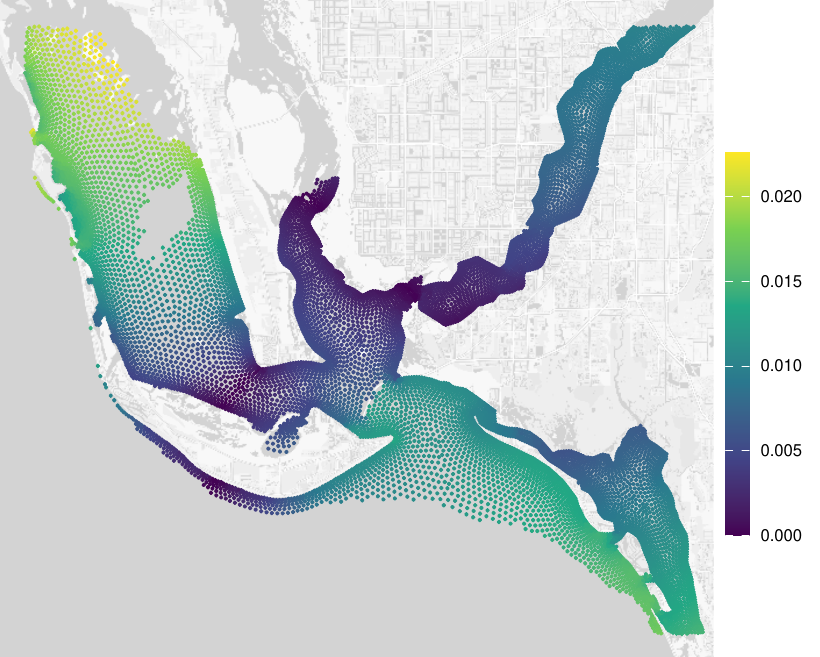}
        \caption{$|\bfk_{2,2}|$.}
    \end{subfigure}
    \vspace{0.5cm}
    
    \caption{Spatial maps of RMSPE and principal component direction under the NONSEP model. Panels (a) and (b) show the RMSPE based on the first PC and first two PCs, respectively. Panel (c) shows the absolute value of the second principal component direction at fidelity level 2. }
    \label{fig: ss_rmspe_pc}
\end{figure}

In the NONSEP model, we categorize the errors in predicting the outputs into two different types: (a) reconstruction error (RE) due to the choice of number of PCs and (b) the model (or emulator) error in predicting the PC weights. The smallest reconstruction error is obtained using a basis representation constructed by performing PCA simultaneously on both the training and testing data and will be referred to as the \textit{minimum reconstruction error} (MRE) in Table~\ref{tab: ss_rmspe_nonsep}. To assess how the fixed basis vectors in \eqref{def: basis_rep}, which are constructed using only training data, influence the total prediction error, the MRE can be compared to the reconstruction error for a reduced-rank representation of the test outputs based on $\bfk_{2,1},\ldots, \bfk_{2,p_{2}}$ in \eqref{def: basis_rep}. We refer to the second reconstruction error as the \textit{testing reconstruction error} (TRE). Both MRE and TRE are computed by measuring the mean square prediction error (MSPE) from projecting a reduced-rank representation of the test outputs back to the original output space. The last two columns of Table \ref{tab: ss_rmspe_nonsep} show the prediction error for the NONSEP emulator (cokriging) and a NONSEP model estimated using only the level-2 weights (kriging). Comparing the MRE and TRE to the cokriging prediction error in Table \ref{tab: ss_rmspe_nonsep}, the choice of basis functions explains the smallest percentage of the total prediction error of the three error sources. The autoregressive structure for the random weights in \eqref{mod: dist_wts} reduces the MSPE for the cokriging emulator with seven PCs by more than $50\%$ compared to the kriging model. Improvements in the predictive performance of the NONSEP emulator for more than seven PCs is negligible. Figure~\ref{fig: ss_rmspe_pc} shows how the RMSPE for the NONSEP emulator with one PC changes after a second PC is added to the model. The locations with the largest reduction in RMSPE after a second PC is added to the model are closely related to the locations with the largest absolute elements of the second principal component direction, $\bfk_{2,2}$, in Equation~\eqref{def: basis_rep}; see Panel (c) of Figure~\ref{fig: ss_rmspe_pc}.

\begin{table}[ht!] 
\centering
\caption{Principal component analysis diagnostics and prediction performance under NONSEP cokriging and NONSEP kriging models as a function of number of PCs. NONSEP cokriging is performed based on outputs at low-fidelity and high-fidelity. NONSEP kriging is performed based on outputs at high-fidelity only. MRE = minimum reconstruction error. TRE = testing reconstruction error.}
\label{tab: ss_rmspe_nonsep}
{\resizebox{1.0\textwidth}{!}{%
\setlength{\tabcolsep}{1.5em}
\begin{tabular}{ccccccc} 
\toprule
  & \multicolumn{2}{c}{$\%$ Variance Explained} & \multicolumn{4}{c}{$10^{2} \times $ MSPE} \\
\cmidrule(lr){2-3} \cmidrule(lr){4-7}
\multirow{2}{*}{\# PCs} &
\multirow{2}{*}{ADCIRC} &
\multirow{2}{*}{ADCIRC + SWAN} &
\multirow{2}{*}{MRE} &
\multirow{2}{*}{TRE} &
\multicolumn{2}{c}{NONSEP} \\
\cline{6-7}
 &  &  &  &  & cokriging & kriging \\
  \hline
  1 & 91.26 & 92.60 & 6.83 & 7.00 & 7.45 & 8.35 \\ 
    2 & 95.93 & 96.87 & 2.64 & 2.79 & 3.33 & 4.46 \\ 
    3 & 97.50 & 98.20 & 1.55 & 1.96 & 2.57 & 3.73 \\ 
    4 & 98.60 & 99.07 & 0.91 & 1.00 & 1.62 & 2.85 \\ 
    5 & 99.32 & 99.50 & 0.44 & 0.49 & 1.17 & 2.50 \\ 
    6 & 99.53 & 99.67 & 0.32 & 0.37 & 1.08 & 2.43 \\ 
    7 & 99.65 & 99.76 & 0.27 & 0.33 & 1.04 & 2.40 \\ 
   \hline
\end{tabular}
}}
\end{table}

\section{Discussion} \label{sec: discussion}

In this paper, we have developed an autoregressive cokriging framework with two multifidelity emulators for modeling computer models with high-dimensional spatial outputs. To effectively account for the cross-covariance in the output space, we have constructed a separable model and nonseparable model. In the SEP ARCokrig emulator, we model the cross-covariance structure as a Kronecker  product of the input correlation and the output cross-covariance. To estimate the cross-covariance matrix, we reformulate the problem as a series of successive autoregressive regressions, which allows us to interpret the choice of prior specification and enables fast shrinkage estimation via  a sparse Cholesky prior, even for high-dimensional outputs. In the NONSEP ARCokrig emulator, we represent the outputs at each fidelity level as a linear combination of basis functions and random weights. The random weights are modeled using an autoregressive cokriging structure across fidelity levels to enforce the rank ordering of output accuracy. 

We have derived closed-form formulas for posterior predictive distributions up to input correlation parameters for the two multivariate GP emulators. Markov chain Monte Carlo algorithms are required only to sample from the posterior distributions of the input correlation parameters, and can be implemented separately for each fidelity level. This is similar to fitting independent multivariate GP models without accounting for the cross-covariance between outputs. The proposed framework thus enables fast Bayesian estimation and prediction. We demonstrate the performance of the proposed emulators in a testbed example and a real application of storm surges. We consistently find that the two emulators with more complex cross-covariance structures can better quantify uncertainties for aggregated quantities over the output space without incurring extra computational burden. The improved uncertainty quantification at aggregated spatial resolutions of the two proposed emulators would help practitioners in coastal flood hazard studies perform risk assessment. 

The proposed framework exploits the practical implications of Markov property via autoregressive structures, separable structure, and nonseparable structure for modeling computer model outputs. These modeling choices have been empirically verified in the numerical example and storm surge application through extensive diagnostics. The Markov property based autoregressive cokriging is suitable to link computer models whose outputs exhibit a natural ordering in terms of accuracy. The modeling choice between a separable structure and a nonseparable structure has received lots of attentions in the literature. Our findings are consistent with prior work \citep{Fricker2013} that for highly rough outputs such as in the testbed example, a separable cross-covariance may give better predictive performance than a PCA-based nonseparable structure. However, the SEP emulator may exhibit slight undercoverage due to its assumption of a constant cross-covariance across the input space. In the storm surge application, the NONSEP emulator achieves superior predictive performance to the SEP emulator.

The two proposed emulators can be further explored. In the SEP ARCokrig model, the selection of neighbors is informed by the smoothness of outputs over the output space. We currently use the maximin design to sort spatial locations and then choose the number of neighbors to define the sparse Cholesky prior. Since there is no parametric assumption imposed, unlike Vecchia's approximation in the purely spatial setting, further investigation is needed to study how the choice of neighbor sets affects shrinkage estimation of the output cross-covariance. In the NONSEP autoregressive cokriging, we use exploratory data analysis tools to specify which low-fidelity PC scores should be included to model the PC scores at high-fidelity via the autoregressive cokriging model. The procedure could be fully automated by employing variable selection methods, such as spike-and-slab priors on the regression coefficients.

There are several potential future research directions. First, in many physical systems, the outputs could be generated at different spatial resolutions. Building a multivariate emulator would need to account for the resolution difference. This could be accounted for by adopting the change of support strategy \citep{Ma2019}. Second, when there are more than two multifidelity computer models, the SEP emulator can be generalized in a straightforward manner; in contrast, there may be further challenges in linking  PC scores at multiple fidelity levels via the autoregressive cokriging framework. The selection of PC scores in the autoregressive cokriging model needs to be treated carefully via techniques such as Bayesian variable selection.

\section*{Supplement Material}

The Supplement Material consists of: (i) technical details and additional results (.pdf file) and (ii) R code to reproduce the numerical results in the example and application (.zip file). 


\begin{singlespace}
\bibliographystyle{apalike}
\setlength{\bibsep}{5pt}
\bibliography{main}
\end{singlespace}

\clearpage\pagebreak\newpage
\thispagestyle{empty}
\begin{center}
{\LARGE{\bf Supplementary Material: {Multivariate Gaussian process emulation for multifidelity computer models with high-dimensional spatial outputs}} } 
\vspace{1cm}
\\ \LARGE {\bf by}
\end{center}

\vskip 1cm
\baselineskip=15pt
\begin{center}
  Cyrus S. McCrimmon and Pulong Ma\\
  {\it Department of Statistics, Iowa State University}\\
  \hskip 5mm\\
\end{center}

\setcounter{equation}{0}
\setcounter{page}{0}
\setcounter{table}{0}
\setcounter{section}{0}
\setcounter{figure}{0}
\clearpage\pagebreak\newpage

\counterwithout{equation}{section}

\renewcommand{\theequation}{S.\arabic{equation}}
\renewcommand{\thesection}{S.\arabic{section}}
\renewcommand{\thesubsection}{S.\arabic{section}.\arabic{subsection}}
\renewcommand{\thepage}{S.\arabic{page}}
\renewcommand{\thetable}{S.\arabic{table}}
\renewcommand{\thefigure}{S.\arabic{figure}}

The Supplementary Material is organized as follows. Section \ref{supp: sup_tech} contains derivations on the technical results from Section \ref{sec: model formulation} of the main paper including detailed derivation for \textbf{Result}~\ref{res: MN_decom} (see Sections \ref{supp: MN_MCD}), Proposition \ref{prop: post_theta} (see Section~\ref{supp: deriv_post_details}),  Proposition \ref{prop: S_pred_dist} (see Section~\ref{supp: post_pred_details}),  and the posterior distribution and posterior predictive distributions of the random weights in Section \ref{sec: nonsep cokriging} (see Section \ref{supp: sup_nonsep}). Section~\ref{sec: sup_sim_diag} includes additional details for the testbed example. Section~\ref{supp: storm surge} contains addition details for the storm surge application.

\section{Technical Results} \label{supp: sup_tech}

\subsection{Derivation for Result 2.1} \label{supp: MN_MCD}

In this section, we provide a derivation of \textbf{Result}~\ref{res: MN_decom}, extending the reparameterization in \cite{pourahmadi1999joint} to a matrix normal random variable. To simplify the notation, we consider the case where the mean is $\bfzero$. At the end of the section, we describe how the result can be extended to a matrix normal random variable with a non-zero mean. Suppose $\bfZ \sim \MN_{n\times N}(\bfzero, \bfR, \bfSigma)$, where $\bfR$ is an $n \times n$ correlation matrix and $\bfSigma$ is an $N \times N$ cross-covariance matrix. Both $\bfR$ and $\bfSigma$ are positive definite. Consider the problem of finding $\bfb_{j}:= (b_{j,1}, \ldots, b_{j, j-1})^{\intercal}$ to minimize \[E[(\bfZ_{j} - \bfZ_{1:j-1}\bfb_{j})^{\intercal} (\bfZ_{j} - \bfZ_{1:j-1}\bfb_{j})],\] where $\bfZ_{j}$ and $\bfZ_{1:j-1}$ are the submatrices of $\bfZ$ containing the $j^{\text{th}}$ and $1, \ldots, j-1$ columns of $\bfZ$, respectively. Rewriting the quadratic form, the previous expectation is equivalent to 

\[\begin{bmatrix}
     -\bfb_j^{\intercal} & 1
\end{bmatrix} E(\bfZ_{1:j}^{\intercal}\bfZ_{1:j}) \begin{bmatrix}
    -\bfb_j\\
    1
\end{bmatrix},\]

\noindent which, from Theorem 2.3.5 in \cite{Gupta1999}, is 
\begin{align} \label{def: quad_form_bj}
    n \begin{bmatrix}
    1 & -\bfb_j^{\intercal} 
\end{bmatrix} \bfSigma_j \begin{bmatrix}
    1\\
    -\bfb_j
\end{bmatrix},
\end{align}
where $\bfSigma_{j}$ is a $j \times j$ matrix comprised of the first $j$ rows and columns of $\bfSigma$. Setting the partial derivative of expression \eqref{def: quad_form_bj} with respect to $\bfb_j$ equal to $0$ and solving for $\bfb_j$ yields the solution $ \bfSigma^{-1}_{j-1}\bfsigma_j$, where $\bfSigma_{j-1}$ and $\bfsigma_j$ are the submatrices of $\bfSigma$ containing the first $j-1$ rows and columns and the $j-1$ elements in row $j$, respectively. Let $\bfa_j := \bfSigma^{-1}_{j-1}\bfsigma_j$. Then $\text{Var}(\bfZ_j - \bfZ_{1:j-1}\bfa_j) = d_j\bfR$, where $ d_j =\sigma_{j,j} -\bfsigma^{\intercal}_{j}\bfSigma^{-1}_{j}\bfsigma_{j}$, and $\sigma_{j,j}$ is the element in row $j$ and column $j$ of $\bfSigma$. From Theorem 2.3.12 in \cite{Gupta1999}, the conditional distribution of \[\bfZ_{j} | \bfZ_{1:j-1} \sim \MN_{n, 1} (\bfZ_{1:j-1}\bfa_j, \bfR, d_{j}).\]

The random vectors $\bfZ_{j} - \bfZ_{1:j-1}\bfa_{j}$ and $\bfZ_{i} - \bfZ_{1:i-1}\bfa_{i}$ are independent for any integers $1 \leq i < j \leq N$; see \cite{pourahmadi1999joint} or  Theorem 2.3.14 in \cite{Gupta1999}. Let $\bfA$ be an $N \times N$ lower-triangular matrix with zeroes along the diagonal and $(a_{j,1}, \ldots, a_{j, j-1}) = \bfa_{j}^{\top}$. Additionally, let $\bfD$ be a diagonal matrix with elements $d_1, \ldots, d_N$. Using independence between $\bfZ_{j} - \bfZ_{1:j-1}\bfa_{j}$ and $\bfZ_{i} - \bfZ_{1:i-1}\bfa_{i}$ for any integers $1 \leq i < j \leq N$ and the fact that $\text{Var}(\bfZ_{j} - \bfZ_{1:j-1}\bfa_{j}) = d_{j}\bfR$,

\[\bfZ(\bfI - \bfA)^{\top} \overset{\text{d}}{=}\MN_{n,N}(\bfzero, \bfR, \bfD).\] From Theorem 2.3.10 in \cite{Gupta1999}, 

\[\bfZ \overset{\text{d}}{=} \MN_{n,N}(\bfzero, \bfR, (\bfI- \bfA)^{-1}\bfD(\bfI -\bfA)^{-\top}),\]
meaning $\bfOmega := \bfSigma^{-1} = (\bfI - \bfA)^{\intercal} \bfD^{-1} (\bfI-\bfA)$. Using the parameterization $\bfSigma^{-1} = (\bfI - \bfA)^{\intercal} \bfD^{-1} (\bfI-\bfA)$ and the definitions of $\bfa_{j}$ and $d_{j}$, the density of $\bfZ$ can be written as
\begin{align} \label{def: cond_dist_likelihood}
    p(\bfZ) &= p(\bfZ_{1})\prod_{j=2}^{N}p(\bfZ_{j} | \bfZ_{1:j-1}),
\end{align}
where $\bfZ_{j} | \bfZ_{1:j-1} \sim \MN_{n,1}(\bfZ_{1:j-1}\bfa_{j}, \bfR, d_{j})$. If $\bfZ \sim \MN_{n,N}(\bfM, \bfR, \bfSigma)$, the same solution to the minimization problem at the start of the section can be obtained after subtracting $\bfM$ from $\bfZ$. Then 
\[(\bfZ -\bfM)(\bfI -\bfA)^{\intercal} \overset{\text{d}}{=} \MN_{n,N}(\bfzero, \bfR, \bfD),
\] 
which simplifies to $\bfZ \sim \MN_{n,N}(\bfM, \bfR, (\bfI -\bfA)^{-1}\bfD(\bfI-\bfA)^{-\top})$. Finally, from Theorem 2.3.12 in \cite{Gupta1999},

\[\bfZ_{j} | \bfZ_{1:j-1} \sim \MN_{n,1}(\bfM_j + (\bfZ_{1:j-1}-\bfM_{1:j-1})\bfa_{j}, \bfR, d_{j}).\]

\subsection{Derivation of Proposition~\ref{prop: post_theta}} \label{supp: deriv_post_details}

In this section, we derive the density of $p(\bftheta | \yObs)$ defined in Proposition 
\ref{prop: post_theta}. The joint posterior distribution of $\bfbeta, \bfA, \bfD, \bftheta$ is proportional to 
\begin{align} \label{eqn: full cond} 
\begin{split}
p(\bfbeta, \bfA, \bfD, \bftheta \mid\bfy^{\mathcal{D}}) &\propto \prod_{t=1}^m \biggr\{|(\textbf{I} - \textbf{A}_{t})^{\intercal}\bfD_{t}^{-1}(\bfI -\bfA_{t})|^{\frac{n_t}{2}
} |\bfR_{t}|^{\frac{-N}{2}}\\
&\quad \times \exp \left \{ - \frac{1}{2} \text{tr} \biggr((\textbf{I} - \textbf{A}_{t})^{\intercal}\bfD_{t}^{-1}(\bfI -\bfA_{t})(\bfY_{t} - \bfF_{t}\bfB_{t})^{\intercal}\bfR_{t}^{-1}(\bfY_{t} - \bfF_{t}\bfB_{t}) \biggr)  \right\}\\
& \quad \times \prod_{l=1}^{d} \frac{2}{\pi(1+(\frac{\theta_{t, \ell}}{q_{t,\ell}})^{2})} \mathds{1}\{\theta_{t,l} >0\} \times d_{t,1}^{-\frac{\eta_t}{2} -1} \exp\left(\frac{-\lambda_{t}}{2d_{t,1}}\right)\\
& \quad \times \prod_{j=2}^{N} (\tau^{2}_{t})^{\frac{-\Cjcard}{2}}(d_{t,j})^{\frac{-\Cjcard}{2}}\exp\left\{-\frac{1}{2\tau^{2}_{t} d_{t,j}} \bfa_{t,j}^{(\Cjset)^\intercal} \bfa_{t,j}^{(\Cjset)}\right\}
d_{t,j}^{-\frac{\eta_t}{2} -1} \exp\left(\frac{-\lambda_{t}}{2d_{t,j}}\right) \biggr\}.
\end{split} 
\end{align}
\noindent After completing the square, 
\begin{align} \label{def: beta_cs}
    &\exp \left \{ - \frac{1}{2} \text{tr} \biggr(\bfOmega_{t}(\bfY_{t} - \bfF_{t}\bfB_{t})^{\intercal}\bfR_{t}^{-1}(\bfY_{t} - \bfF_{t}\bfB_{t}) \biggr)  \right\} \notag\\
    &=\exp \left \{ - \frac{1}{2} \text{tr} \biggr(\bfOmega_{t}(\bfY_{t} - \bfF_{t}\hat{\bfB}_{t})^{\intercal}\bfR_{t}^{-1}(\bfY_{t} - \bfF_{t}\hat{\bfB}_{t}) \biggr)  \right\}\notag\\
    &\times \exp \left \{ - \frac{1}{2} \text{tr} \biggr(\bfOmega_{t}(\bfbeta_{t} - \hat{\bfbeta}_{t})^{\intercal}(\bfH_{t}^{\intercal}\bfR_{t}^{-1}\bfH_{t})(\bfbeta_{t} - \hat{\bfbeta}_{t}) \biggr)  \right\}
\end{align}
where $\hat{\bfbeta}_{t}$ and $\hat{\bfB}_{t}$ are defined in Section \ref{sec: posterior inference}. Since the second term in Expression \eqref{def: beta_cs} is equivalent to the kernel of a matrix normal distribution, the posterior distribution \eqref{eqn: full cond} can be rewritten as
\begin{align} \label{eqn: before_beta_int_sup}
p(\bfA, \bfD, \bftheta | \bfy^{\mathcal{D}}) &\propto \prod_{t=1}^m \biggr\{|\bfOmega_{t}|^{\frac{n_t}{2}} |\bfR_{t}|^{\frac{-N}{2}} |\bfH^{\intercal}_{t}\bfR_{t}^{-1}\bfH_{t}|^{\frac{-N}{2}} \notag \\  
& \times \exp \left \{ - \frac{1}{2} \text{tr} \biggr(\bfOmega_{t}(\bfY_{t} - \bfF_{t}\hat{\bfB}_{t})^{\intercal}\bfR_{t}^{-1}(\bfY_{t} - \bfF_{t}\hat{\bfB}_{t}) \biggr)  \right\}\notag\\
& \quad \times \prod_{l=1}^{d} \frac{2}{\pi(1+(\frac{\theta_{t, \ell}}{q_{t,\ell}})^{2})}\times d_{t,1}^{-\frac{\eta_t}{2} -1} \exp\left(\frac{-\lambda_{t}}{2d_{t,1}}\right) \notag\\
& \quad \times \prod_{j=2}^{N} (\tau^{2}_{t})^{\frac{-\Cjcard}{2}}(d_{t,j})^{\frac{-\Cjcard}{2}}\exp\left\{-\frac{1}{2\tau^{2}_{t}d_{t,j}}\bfa_{t,j}^{(\Cjset)^\intercal}\bfa_{t,j}^{(\Cjset)}\right\} d_{t,j}^{-\frac{\eta_t}{2} -1} \exp\left(\frac{-\lambda_{t}}{2d_{t,j}}\right)\biggr\}
\end{align}
where
\begin{align} \label{def: fc_beta}
    p(\bfbeta|  \bfA, \bfD, \bftheta, \bfy^{\mathcal{D}}) \propto \prod_{t=1}^{m}\MN_{q_{t},N}(\hat{\bfbeta}_{t}, (\bfH^{\intercal}_{t}\bfR_{t}^{-1}\bfH_{t})^{-1}, \bfOmega_{t}^{-1}).
\end{align}
Integrating $\bfbeta$ out of expression \eqref{eqn: before_beta_int_sup} yields the distribution of $p(\bfA, \bfD, \bftheta | \bfy^{\mathcal{D}})$ defined as
\begin{align} \label{eqn: after_beta_int}
p(\bfA, \bfD, \bftheta | \bfy^{\mathcal{D}}) &\propto \prod_{t=1}^m \biggr\{|\bfOmega_{t}|^{\frac{n_t}{2}} |\bfR_{t}|^{\frac{-N}{2}} |\bfH^{\intercal}_{t}\bfR_{t}^{-1}\bfH_{t}|^{\frac{-N}{2}} \notag\\
&\quad \times \exp \left \{ - \frac{1}{2} \text{tr} \biggr(\bfOmega_{t}(\bfY_{t} - \bfF_{t}\hat{\bfB}_{t})^{\intercal}\bfR_{t}^{-1}(\bfY_{t} - \bfF_{t}\hat{\bfB}_{t}) \biggr)  \right\}\notag\\
& \quad \times \prod_{l=1}^{d} \frac{2}{\pi(1+(\frac{\theta_{t, \ell}}{q_{t,\ell}})^{2})}\times d_{t,1}^{-\frac{\eta_t}{2} -1} \exp\left(\frac{-\lambda_{t}}{2d_{t,1}}\right) \notag\\
& \quad \times \prod_{j=2}^{N} (\tau^{2}_{t})^{\frac{-\Cjcard}{2}}(d_{t,j})^{\frac{-\Cjcard}{2}}\exp\left\{-\frac{1}{2\tau^{2}_{t}d_{t,j}}\bfa_{t,j}^{(\Cjset)^\intercal}\bfa_{t,j}^{(\Cjset)}\right\} d_{t,j}^{-\frac{\eta_t}{2} -1} \exp\left(\frac{-\lambda_{t}}{2d_{t,j}}\right)\biggr\}.
\end{align}
Since $(\bfI-\bfA_{t})$ and $\bfD_{t}$ are a lower triangular and diagonal matrix, respectively, \[|\bfOmega_{t}| =|(\textbf{I} - \textbf{A}_{t})^{\intercal}\bfD_{t}^{-1}(\bfI -\bfA_{t})| = \prod_{j=1}^{N}d^{\frac{-1}{2}}_{t,j}, \] where $d_{t,j}$ is the $j^{\text{th}}$ element along the diagonal of $\bfD_{t}$. Recall that $\bfa_{j}^{(\Cjset)}$ is a column vector comprised of the non-zero elements in row $j$ of $\bfA_{t}$. Using the results in Section~\ref{supp: MN_MCD},
\begin{align} \label{eqn: dens_fact_S}
    & |\bfOmega_{t}|^{\frac{n_{t}}{2}}\exp \left \{ - \frac{1}{2} \text{tr} \biggr(\bfOmega_{t}(\bfY_{t} - \bfF_{t}\hat{\bfB}_{t})^{\intercal}\bfR_{t}^{-1}(\bfY_{t} - \bfF_{t}\hat{\bfB}_{t}) \biggr)  \right\} \notag\\
    &= \prod_{j=1}^{N}d^{\frac{-n_{t}}{2}}_{t,j}\exp \left \{ - \frac{1}{2} \text{tr} \biggr(\bfOmega_{t}\bfS_{t}^{\intercal}\bfS_{t} \biggr)  \right\}\notag\\
    &=\prod_{j=1}^{N}d^{\frac{-n_{t}}{2}}_{t,j}\exp \left \{ - \frac{1}{2d_{t,j}} (\bfS_{t,j}-\bfS_{t,\Cjset}\bfa^{(\Cjset)}_{t,j})^{\intercal}(\bfS_{t,j}-\bfS_{t,\Cjset}\bfa^{(\Cjset)}_{t,j})   \right\}
\end{align}
where $\bfS_{t}$ is defined in Proposition \ref{prop: post_theta}. Recall that $\bfS_{t,j}$ and $\bfS_{t,\Cjset}$ are the submatrices of $\bfS_t$ that include the $j^{\text{th}}$ column and each column in the set $\Cjset$, respectively. Using the factorization in equation \eqref{eqn: dens_fact_S} and expression \eqref{eqn: after_beta_int}, 
\begin{align*}
    p(\bfA, \bfD |\bftheta,\yObs) &\propto \prod_{t=1}^{m}\{ d^{\frac{-(n_{t}-q_{t})}{2}}_{t,1} \exp \{ - \frac{1}{2d_{t,1}} \bfS^{\intercal}_{t,1}\bfS_{t,1}\}d_{t,1}^{-\frac{\eta_{t}}{2} -1} \exp(\frac{-\lambda_{t}}{2d_{t,1}})\notag\\
    &\quad \times  \prod_{j=2}^{N} d^{\frac{-(n_{t}-q_{t})}{2}}_{t,j} \exp \left \{ - \frac{1}{2d_{t,j}} (\bfS_{t,j}-\bfS_{t,\Cjset}\bfa^{(\Cjset)}_{t,j})^{\intercal}(\bfS_{t,j}-\bfS_{t,\Cjset}\bfa^{(\Cjset)}_{t,j})   \right\}\notag\\
    & \quad \times  d_{t,j}^{\frac{-\Cjcard}{2}}\exp\{-\frac{1}{2\tau^{2}_{t}d_{t,j}}\bfa_{t,j}^{(\Cjset)^\intercal}\bfa_{t,j}^{(\Cjset)}\} d_{t,j}^{-\frac{\eta_{t}}{2} -1} \exp(\frac{-\lambda_{t}}{2d_{t,j}}) \}
\end{align*}
which, after completing the square, is proportional to
\begin{align} \label{def: sup_abi_factorization}
    p(\bfA, \bfD | \bftheta, \yObs)  &\propto \prod_{t=1}^{m}
    p(d_{t,1} | \bftheta_t, \yObs)\prod_{j=2}^{N} p(\bfa_{t,j}^{(\Cjset)} | d_{t,j}, \bftheta_t, \yObs)p(d_{t,j}| \bftheta_t, \yObs)\notag\\
    &\hat{d}_{t,1}^{-\frac{\nu_{t}}{2}} \prod_{j=2}^{N}|\hat{\bfV}_{t,j}|^{\frac{1}{2}}\hat{d}_{t,j}^{\frac{-\nu_{t}}{2}},
\end{align}
with
\begin{align}
    \bfa_{t,j}^{(\Cjset)} | d_{t,j}, \bftheta_t, \yObs &\sim \MVN_{\Cjcard}(\hat{\bfa}_{t,j}^{(\Cjset)}, d_{t,j}\hat{\bfV}_{t,j}) \label{def: post_a}\\
    d_{t,j}| \bftheta_t, \yObs &\sim\text{IG}(\frac{\nu_t}{2},\frac{\hat{d}_{t,j}}{2}), \label{def: post_d}
\end{align}
for $\hat{\bfa}_{t,j}^{(\Cjset)}$, $\hat{\bfV}_{t,j}$, $\hat{d}_{t,j}$ and $\nu_t$ defined in Proposition \ref{prop: post_theta}. Using expression \eqref{def: sup_abi_factorization} to integrate $\bfA$ and $\bfD$ out of the density in \eqref{eqn: after_beta_int} yields the expression for $ p(\bftheta | \yObs)$ defined in Proposition \ref{prop: post_theta}.

\subsection{Derivation in Proposition \ref{prop: S_pred_dist}} \label{supp: post_pred_details}

In this section, we derive the distribution of $p(\bfy(\bfx_0) | \bftheta, \yObs)$ defined in Proposition \ref{prop: S_pred_dist}. Using the separable autoregressive model \eqref{eqn: SEP cokriging}, the joint distribution of $\bfy(\bfx_0)$ and $\bfy^{\mathscr{D}}$ given all model parameters is a product of matrix-normal distributions across $m$ fidelity levels:
\begin{align} \label{eqn: joint MN distribution}
\begin{split}
\begin{pmatrix} 
\bfy(\bfx_0) \\
\bfy^{\mathscr{D}} \\
\end{pmatrix} 
\biggr \rvert  \bfbeta, \bfA, \bfD, \bftheta &\sim 
 \prod_{t=1}^m
\mathcal{MN}_{n_t+1, N} \left(   
\begin{pmatrix} 
\bfh^\top_t(\bfx_0) \bfbeta_t + \gamma_{t-1} \bfy_{t-1,1:N}^\top(\bfx_0)  \\
\bfF_{t}\bfB_{t} \\
\end{pmatrix}, \right. \\
&\quad\quad \left.
\begin{pmatrix} 
r(\bfx_0, \bfx_0; \bftheta_t) & \bfr_t^\top(\bfx_0)  \\
\bfr_t(\bfx_0)& \bfR_t  \\
\end{pmatrix}, \bfOmega^{-1}_t
\right),
\end{split}
\end{align}
where $\bfr_t(\bfx_0)$ is defined following expression \eqref{def: R_t_star_star}. Since the distribution in \eqref{eqn: joint MN distribution} features a product of matrix normal distributions, standard calculations yield the predictive distribution  (\citeauthor{Conti2010}, \citeyear{Conti2010}; \citeauthor{Gupta1999}, \citeyear{Gupta1999}, Theorem 2.3.12)
\begin{align} \label{dist: y_x0 cond all_beforebeta}
    \bfy(\bfx_0) | \bfbeta, \bfA, \bfD, \bftheta, \bfy^{\mathscr{D}} & \sim \prod_{t=1}^{m}\MN_{1,N}(\bfM_{t, *}, R_{t, *}, \bfOmega^{-1}_{t}),
\end{align}
\noindent where
\begin{align*}
    \bfM_{t,*} &= \begin{cases}
        h^{\intercal}_{1}(\bfx_{0})\bfbeta_{1}  + \bfr^{\intercal}_{1}(\bfx_{0})\bfR_{1}^{-1}(\bfy_{1} - \bfF_{1}\bfB_{1})&\text{ if } t=1\\
        h^{\intercal}_{t}(\bfx_{0})\bfbeta_{t} + \gamma_{t-1}\bfy_{t-1}(\bfx_{0}) + \bfr^{\intercal}_{t}(\bfx_{0})\bfR_{t}^{-1}(\bfy_{t} - \bfF_{t}\bfB_{t})&\text{ if } t>1
    \end{cases} \\
    R_{t,*}&= 1 - \bfr^{\intercal}_{t}(\bfx_{0})\bfR_{t}^{-1}\bfr_{t}(\bfx_{0}) . 
\end{align*}
\noindent A similar strategy is exploited to integrate $\bfbeta$ out of the distribution in \eqref{dist: y_x0 cond all_beforebeta}. Since the distribution of $\bfbeta| \bfA, \bfD, \bftheta, \yObs$ defined in \eqref{def: fc_beta} is proportional to a product of matrix normal densities,  the density of $\bfy_{t}(\bfx_{0}) | \bfy_{t-1}(\bfx_{0}), \bfA, \bfD, \bftheta, \yObs$ will be proportional to a matrix normal density. To obtain the moments for the distribution, note that for $t>1$,
\begin{align} \label{def: sup_M_t_star_star}
     \Expec[\bfy_{t}(\bfx_{0})| \bfy_{t-1}(\bfx_{0})]&=\Expec_{\bfbeta_{t}}[\Expec[\bfy_{t}(\bfx_{0})| \bfy_{t-1}(\bfx_{0}), \bfbeta_{t}]] \notag\\
     &= \Expec_{\bfbeta_{t}}[\bfM_{t,*}]\notag\\
     &= h^{\intercal}_{t}(\bfx_{0})\hat{\bfbeta}_{t} + \gamma_{t-1}\bfy_{t-1}(\bfx_{0}) + \bfr^{\intercal}_{t}(\bfx_{0})\bfR_{t}^{-1}(\bfy_{t} - \bfF_{t}\hat{\bfB}_{t}),
\end{align}
\noindent which is equal to $\bfM_{t}(\bfx_0)^{\intercal}$ in \eqref{def: M_t_star_star}. A similar calculation provides the expression for $\bfM_{1}(\bfx_0)$. Using the law of total variance,
\begin{align}
    \text{Var}(\bfy_{t}(\bfx_{0}) | \bfy_{t-1}(\bfx_{0})) &= \Expec_{\bfbeta_{t}}[\text{Var}(\bfy_{t}(\bfx_{0}) |\bfy_{t-1}(\bfx_{0}), \bfbeta_{t})] +\text{Var}_{\bfbeta_{t}}(\Expec[\bfy_{t}(\bfx_{0})|\bfy_{t-1}(\bfx_{0}), \bfbeta_{t}])\notag\\
    &= \Expec_{\bfbeta_{t}}[R_{t,*}\otimes\bfOmega_{t}^{-1}] \notag\\
    & + \text{Var}_{\bfbeta_{t}}(h^{\intercal}_{t}(\bfx_{0})\bfbeta_{t} + \gamma_{t-1}\bfy_{t-1}(\bfx_{0}) + \bfr^{\intercal}_{t}(\bfx_{0})\bfR_{t}^{-1}(\bfy_{t} - \bfF_{t}\bfB_{t})), \label{def: var_beta_int}
\end{align}
\noindent where $\otimes$ is the Kronecker product. After applying the definition of $\bfB_{t}$ and dropping constants, the variance term on the right-hand-side of \eqref{def: var_beta_int} is equivalent to \[\text{Var}_{\bfbeta_{t}}(h^{\intercal}_{t}(\bfx_{0})\bfbeta_{t} - \bfr^{\intercal}_{t}(\bfx_{0})\bfR_{t}^{-1}\bfH_{t}\bfbeta_{t}).\] From Theorem 2.3.10 in \cite{Gupta1999} and expression \eqref{def: fc_beta}, 
\begin{align} \label{def: var_beta_int_2}
    &\text{Var}_{\bfbeta_{t}}(h^{\intercal}_{t}(\bfx_{0})\bfbeta_{t} - \bfr^{\intercal}_{t}(\bfx_{0})\bfR_{t}^{-1}\bfH_{t}\bfbeta_{t})\notag\\
    &= (h^{\intercal}_{t}(\bfx_{0})-\bfr^{\intercal}_{t}(\bfx_{0})\bfR_{t}^{-1}\bfH_{t})(\bfH_{t}\bfR_{t}^{-1}\bfH_{t})^{-1}(h^{\intercal}_{t}(\bfx_{0})-\bfr^{\intercal}_{t}(\bfx_{0})\bfR_{t}^{-1}\bfH_{t})^{\intercal}\otimes\bfOmega_{t}^{-1}.
\end{align}
\noindent After combining expressions \eqref{def: var_beta_int} and \eqref{def: var_beta_int_2} and using properties of the Kronecker product, \[\text{Var}(\bfy_{t}(\bfx_{0}) | \bfy_{t-1}(\bfx_{0})) = R_{t}(\bfx_0)\otimes\bfOmega_{t}^{-1},\] where $R_{t}(\bfx_0)$ is defined in \eqref{def: R_t_star_star}. Thus,
\begin{align} \label{dist: y_x0 cond all}
    \bfy(\bfx_0) | \bfA, \bfD, \bftheta, \bfy^{\mathscr{D}} & \sim \prod_{t=1}^{m}\MN_{1,N}(\bfM_{t}(\bfx_0), R_{t}(\bfx_0), \bfOmega^{-1}_{t}).
\end{align}
\noindent Using the factorization in Result \ref{res: MN_decom}, 
\[p(\bfy_t(\bfx_0) | \bfA, \bfD, \bftheta, \yObs) \sim p(\bfy_{t,1} | \bfA, \bfD,\bftheta, \yObs) \prod_{j=2}^{N}p(\bfy_{t,j} | \bfy_{t,\Cjset}, \bfA, \bfD, \bftheta, \yObs),\] where
\[\bfy_{t,j} | \bfy_{t,\Cjset}, \bfA, \bfD, \bftheta, \yObs \sim \MVN_1(\bfM_{t,j}(\bfx_0) + (\bfy_{t,\Cjset} - \bfM_{t,\Cjset})^{\intercal}\bfa_{t,j}^{\Cjset}, d_{t,j}R_{t}(\bfx_0)).\] Then for each fidelity level, the parameters $\bfa_{t,j}^{\Cjset}$ and $d_{t,j}$ can be integrated out using the distributions in \eqref{def: post_a} and \eqref{def: post_d} and standard conjugacy results, yielding the expression for $p(\bfy(\bfx_0) | \bftheta, \yObs)$ in Proposition \ref{prop: S_pred_dist}.

\subsection{Posterior Predictive Distribution in Section \ref{sec: nonsep cokriging}} \label{supp: sup_nonsep}

In this section, we provide expressions for $p(\bftheta_{w} | \wObs)$ and $p(\bfw(\bfx_0) | \bftheta_w, \wObs)$ from Section \ref{sec: nonsep cokriging}. Let $\hat{\bfbeta}_{t-1, \ell}:= (\bfW_{t-1}^{\intercal}\bfR_{t,\ell}^{-1}\bfW_{t-1})^{-1}\bfW_{t-1}^{\intercal}\bfR_{t,\ell}^{-1}\bfw_{t, \ell}$ be the generalized least-squares estimator of $\bfbeta_{t-1, \ell}$. Using standard conjugacy results, it can be shown that
\begin{align*}
    p(\bftheta_{w} | \wObs)  \propto & \prod_{\ell=1}^{p_1} |\bfR_{1, \ell}|^{\frac{-1}{2}} (\bfw_{1,\ell}^{\intercal}\bfR_{1,\ell}^{-1}\bfw_{1,\ell})^{\frac{-n_1}{2}}\pi(\bftheta_{1,\ell}) \\
    & \prod_{t=2}^{m}\prod_{\ell=1}^{p_t} \biggl\{ |\bfR_{t, \ell}|^{\frac{-1}{2}} |\bfW_{t-1}^{\intercal}\bfR_{t,\ell}^{-1}\bfW_{t-1}|^{\frac{-1}{2}}\\
    &((\bfw_{t,\ell} - \bfW_{t-1}\hat{\bfbeta}_{t-1, \ell})^{\intercal}\bfR_{t,\ell}^{-1}(\bfw_{t,\ell} - \bfW_{t-1}\hat{\bfbeta}_{t-1, \ell}))^{\frac{-(n_t -p_{t-1})}{2}}\pi(\bftheta_{t,\ell})\biggr \}
\end{align*}
Let $\bfr_{t,\ell}(\bfx_0):=r(\bfx_0,\mathcal{X}_t; \bftheta_{t,\ell})$ be an $n_t \times 1$ vector of correlations between $\bfx_0$ and the inputs in $\mathcal{X}_t$. The posterior predictive distribution for the random weights is 
\begin{align*}
    p(\bfw(\bfx_0) | \bftheta_w, \wObs) \propto&  \prod_{\ell=1}^{p_1} t_{n_1}(w_{1,\ell}(\bfx_0);m_{1, \ell}(\bfx_0), 
\hat{\sigma}^{2}_{1, \ell}r_{1,\ell}(\bfx_0)) \\
&\prod_{t=2}^{m}\prod_{\ell=1}^{p_m} t_{n_t -p_{t-1}}(w_{t,\ell}(\bfx_0) | w_{t-1,\ell}(\bfx_0); m_{t, \ell}(\bfx_0), 
\hat{\sigma}^{2}_{t, \ell}r_{t,\ell}(\bfx_0)),
\end{align*}
where
\begin{align*}
    m_{1, \ell}(\bfx_0) &= \bfr_{1,\ell}(\bfx_0)^{\intercal}\bfR_{1,\ell}^{-1}\bfw_{1,\ell}, &
    r_{1,\ell}(\bfx_0) &= 1 - \bfr_{1,\ell}(\bfx_0)^{\intercal}\bfR_{1,\ell}^{-1}\bfr_{1,\ell}(\bfx_0), &
    \hat{\sigma}^{2}_{1,\ell} &= n_1^{-1}\bfw_{1,\ell}^{\intercal}\bfR_{1,\ell}^{-1}\bfw_{1,\ell},
\end{align*}
and for $t>1$,
\begin{align*}
    &m_{t, \ell}(\bfx_0) = \bfw_{t-1}(\bfx_0)^{\intercal}\hat{\bfbeta}_{t-1, \ell} +  \bfr_{t,\ell}(\bfx_0)^{\intercal}\bfR_{t,\ell}^{-1}(\bfw_{t,\ell}-\bfW_{t-1}\hat{\bfbeta}_{t-1, \ell}), \\
    &r_{t,\ell}(\bfx_0) = 1 - \bfr_{t,\ell}(\bfx_0)^{\intercal}\bfR_{t,\ell}^{-1}\bfr_{t,\ell}(\bfx_0) \\ 
    &+ (\bfw_{t-1, \ell}(\bfx_0) -\bfr_{t,\ell}(\bfx_0)^{\intercal}\bfR_{t,\ell}^{-1}\bfW_{t-1})^{\intercal} (\bfW_{t-1}^{\intercal}\bfR_{t,\ell}^{-1}\bfW_{t-1})^{-1}(\bfw_{t-1, \ell}(\bfx_0) -\bfr_{t,\ell}(\bfx_0)^{\intercal}\bfR_{t,\ell}^{-1}\bfW_{t-1}),\\
    &\hat{\sigma}^{2}_{1,\ell} = (n_t-p_{t-1})^{-1}(\bfw_{t,\ell} - \bfW_{t-1}\hat{\bfbeta}_{t-1, \ell})^{\intercal}\bfR_{t,\ell}^{-1}(\bfw_{t,\ell} - \bfW_{t-1}\hat{\bfbeta}_{t-1, \ell}).
\end{align*}

\clearpage

\section{Additional Results for the Testbed Example} \label{sec: sup_sim_diag}

In Section \ref{supp: SEP_testbed}, we justify several modeling decisions for the SEP emulator, including the nonstationary cross-covariance matrix, autoregressive structure for outputs at multiple fidelity levels and the choice of hyperparameters for the prior on $\bfA_t$ and $\bfD_t$. In Section \ref{supp: nonsep_testbed}, we describe the implementation details of the NONSEP emulator. In Section \ref{supp: additional visual comparison}, we include visual comparisons of the predictive performance for the SEP and NONSEP emulators.

\subsection{Diagnostics and Implementation Details for the SEP Emulator} \label{supp: SEP_testbed}

\begin{figure}[h!]
    \centering
    \label{fig: EDA testbed}
    \begin{minipage}{\linewidth}
        \begin{subfigure}[b]{0.49\linewidth}
        \centering
        \includegraphics[width=\linewidth]{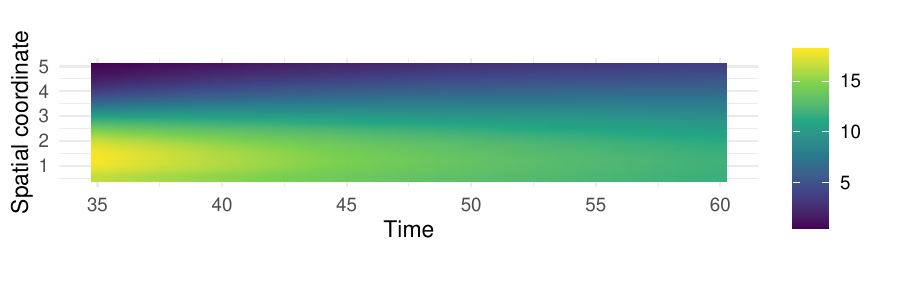}
        \caption{Mean $t=1$.}
    \end{subfigure}
    \hfill
    \begin{subfigure}[b]{0.49\linewidth}
        \centering
        \includegraphics[width=\linewidth]{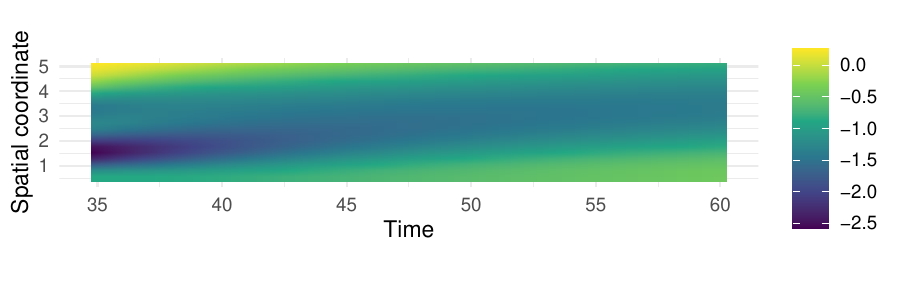}
        \caption{Mean $t=2$.}
        \label{fig: sim_mean_L2}
    \end{subfigure}
    \end{minipage}
    \vspace{-0.2cm}
    \begin{minipage}{\linewidth}
        \begin{subfigure}[b]{0.49\linewidth}
        \centering
        \includegraphics[width=\linewidth]{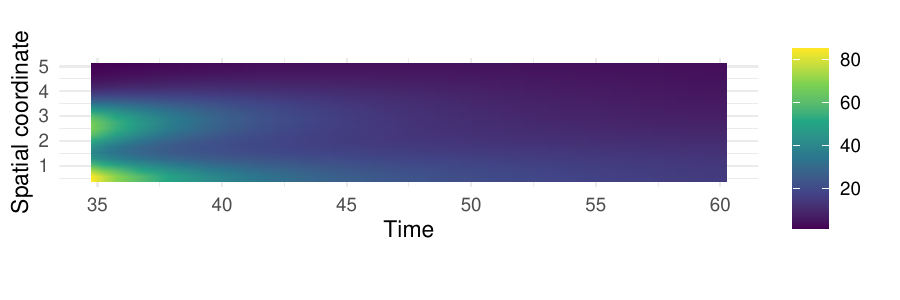}
        \caption{Variance $t=1$.}
    \end{subfigure}
    \hfill
    \begin{subfigure}[b]{0.49\linewidth}
        \centering
        \includegraphics[width=\linewidth]{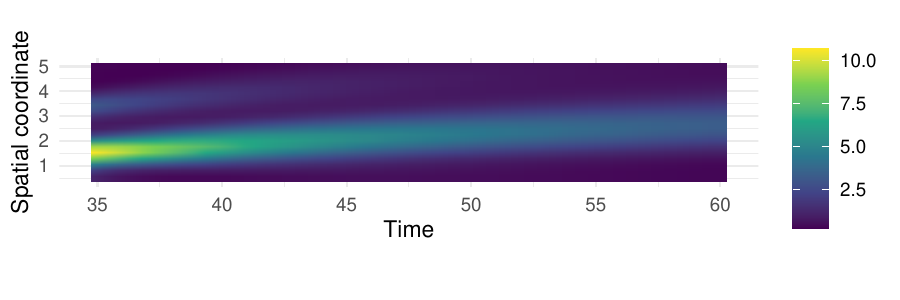}
        \caption{Variance $t=2$.}
    \end{subfigure}
    \end{minipage}
    \vspace{-0.2cm}
    \begin{minipage}{\linewidth}
        \begin{subfigure}[b]{0.49\linewidth}
        \centering
        \includegraphics[width=\linewidth]{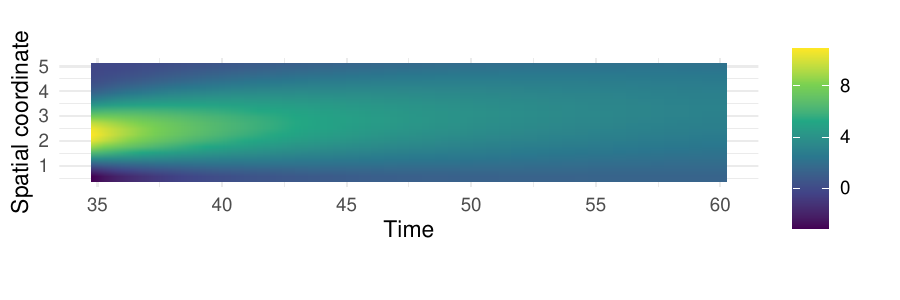}
        \caption{Residual $t=1$.}
    \end{subfigure}
    \hfill
    \begin{subfigure}[b]{0.49\linewidth}
        \centering
        \includegraphics[width=\linewidth]{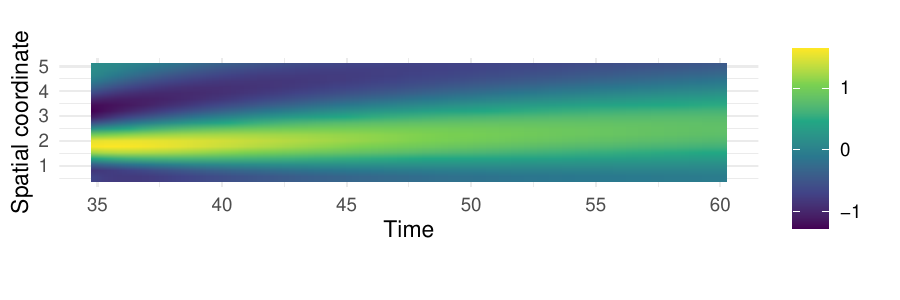}
        \caption{Residual $t=2$.}
    \end{subfigure}
    \end{minipage}

    \caption{Exploratory plots of the simulation in the testbed example. 
    Panels (a) and (b) show the empirical mean for level-1 and level-2 outputs, respectively; panels (c) and (d) show the empirical variance of level-1 and level-2 outputs. Panels (e) and (f) show the residuals at fidelity level 1 and fidelity level 2 after subtracting the mean functions over the input setting $M=12.49, D=0.07, L=  2.03$ and $T=30$.} 
    \label{fig: EDA testbed}
\end{figure}

The first two rows of Figure \ref{fig: EDA testbed} show the empirical mean and variance of the level-1 and level-2 outputs. At $t=2$, we subtracted the level-1 outputs from the level-2 outputs before calculating the mean and variance. At both fidelity levels, the mean and variance vary by space/time location, with the largest variability occurring near time $35$. Based on the difference in variance between panels (c) and (d) of Figure \ref{fig: EDA testbed}, the level-1 outputs help explain a lot of the variability in the level-2 outputs, justifying the autoregressive structure in model \eqref{eqn: SEP cokriging}. After accounting for the level-1 outputs in the mean function for the level-2 outputs and removing the spatial trend, the residuals for nearby locations are still correlated (see Panel (f) of Figure \ref{fig: EDA testbed} for an illustration at a single input).

\begin{figure}[hbt!] 
    \centering
    \begin{subfigure}[b]{0.98\linewidth}
        \includegraphics[
  width=\linewidth,
  trim=0 0.4cm 0 0.2cm,
  clip
]{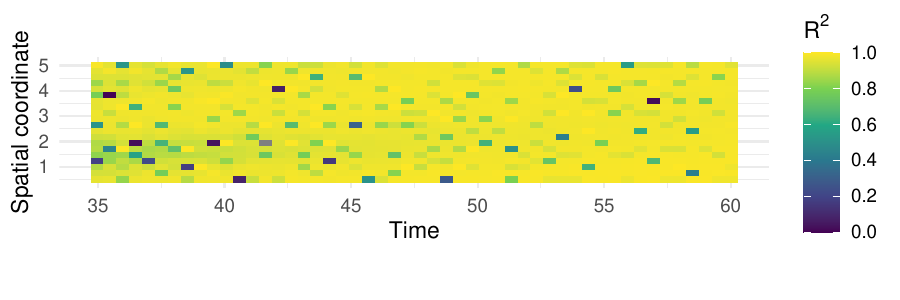}
    \caption{Level 1}
    \end{subfigure}
    \begin{subfigure}[b]{0.98\linewidth}
        \includegraphics[
  width=\linewidth,
  trim=0 0.4cm 0 0.2cm,
  clip
]{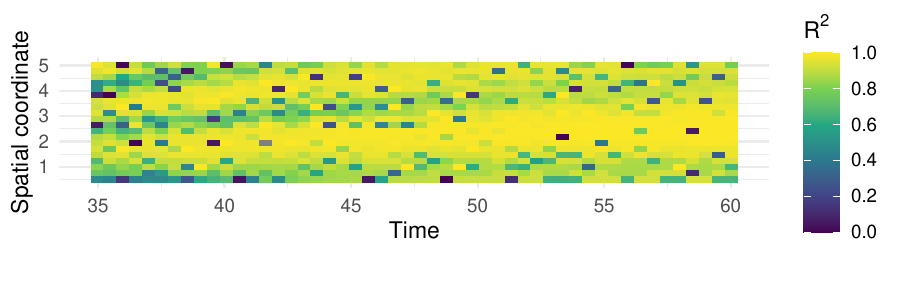}
    \caption{Level 2}
    \end{subfigure}
    \caption{Output variability at each space/time location explained by the nearest neighbor's outputs. Panel (a) shows the coefficient of determination for the centered outputs at a location and its nearest neighbor for the level-1 simulator. Panel (b) shows the coefficient of determination for the centered difference between the level 2 and level 1 outputs.} 
    \label{fig: r_sq_testbed}
\end{figure}

After sorting the spatial locations, we found that a single nearest neighbor explains most of the output variability at many locations. Figure \ref{fig: r_sq_testbed} shows the coefficient of determination from regressing the centered outputs at a location on the centered outputs of its nearest neighbor. The outputs were centered at the empirical mean for each location. The results for level 2 in panel (b) of Figure \ref{fig: r_sq_testbed} were calculated using the difference between the level-2 and level-1 outputs. The smallest values of $R^{2}$ in Figure \ref{fig: r_sq_testbed} tend to occur at earlier space/time locations, where a point's nearest neighbor is typically farther away than it is at later locations. In general, we found that at locations where $R^{2}$ is close to $1$, when the neighbor set includes more than one point, the centered columns of neighbor outputs are nearly linearly dependent, causing numerical problems when evaluating the posterior distribution in \eqref{def: post_theta}.

The results for the SEP emulator in Table~\ref{tab: sim_pred} were obtained after fixing the hyperparameters for the prior on $\bfA_t$ and $\bfD_t$ defined in \eqref{def: prior ad} at $\tau^{2}_t =1, \eta_t=4$ and $\lambda_t=2$ for $t=1,2$. To make predictions, the Metropolis-Hastings algorithm was run for $30,000$ iterations (with a burn-in of $3,000$ samples). The MCMC samples did not show any lack of convergence behaviors. 

The predictive performance of the SEP model is not particularly sensitive to the choice of hyperparameters. We calculated the RMSPE, CVG ($95\%$) and ALCI ($95\%)$ using the MAP estimator of $\bftheta$ for three prior distributions on $d_{t,j}$. Ordered from least to most informative, the hyperparameters were fixed at $(\eta_t, \lambda_t)$= $(0.2, 0.02), (4,2), (4, 0.002)$. The results are shown in Table \ref{tab: sim_pred_hp} with $\tau^{2}_t$ fixed at $1$. The predictive performance at different values of $\tau^{2}_t$ similar to what is presented in Table \ref{tab: sim_pred_hp}. The RMSPE of $\Bar{y}_2(\cdot)$ is omitted from Table \ref{tab: sim_pred_hp} because it is similar for all hyperparameter configurations. The RMSPE hardly changes with $\eta_t$ and $\lambda_t$. The average length of a $95\%$ equal-tail credible interval changes more than the RMSPE for different values of $\eta_t$ and $\lambda_t$, but that variability is small relative to the overall variability of $y_2(\bfx, \bfs)$.

\begin{table}[hbt!]
\centering
\caption{Predictive performance of SEP model for different values of the hyperparameters $\eta_t$ and $\lambda_t$. Predictions were computed using the MAP estimator of $\bftheta$. } 
\label{tab: sim_pred_hp}
\begin{tabular}{ccccccccc}
\toprule
  \multicolumn{4}{c}{Hyperparameters} & \multicolumn{3}{c}{Predictive performance of $y_{2,j}(\cdot)$} & \multicolumn{2}{c}{Predictive performance of $\Bar{y}_2(\cdot)$} \\
\cmidrule(lr){1-4} \cmidrule(lr){5-7} \cmidrule(lr){8-9}
$\eta_1$ & $\lambda_1$ & $\eta_2$ & $\lambda_2$ & RMSPE & CVG $(95\%)$  & ALCI $(95\%)$ & CVG $(95\%)$   &  ALCI $(95\%)$  \\ 
  \hline
\multirow{3}{*}{0.2} & \multirow{3}{*}{0.02} & 0.2 & 0.02 & 0.196 & 96.50 & 0.700 & 93.33 & 0.147 \\ 
  & & 4 & 2 & 0.195 & 96.32 & 0.681 & 93.33 & 0.145 \\ 
  & & 4 & 0.002 & 0.193 & 96.76 & 0.760 & 93.33 & 0.171 \\ 
  \hline
  \multirow{3}{*}{4} & \multirow{3}{*}{2} & 0.2 & 0.02 & 0.196 & 96.52 & 0.701 & 93.33 & 0.150 \\ 
  & & 4 & 2 & 0.195 & 96.37 & 0.682 & 93.33 & 0.148 \\ 
  & & 4 & 0.002 & 0.193 & 96.83 & 0.761 & 93.33 & 0.174 \\ 
  \hline
  \multirow{3}{*}{4} & \multirow{3}{*}{0.002} & 0.2 & 0.02 & 0.195 & 96.59 & 0.704 & 93.33 & 0.152 \\ 
  & & 4 & 2 & 0.195 & 96.44 & 0.685 & 9.333 & 0.150 \\ 
  & & 4 & 0.002 & 0.193 & 96.90 & 0.763 & 93.33 & 0.176 \\  
   \hline
\end{tabular}
\end{table}

\subsection{Diagnostics and Implementation Details for the NONSEP emulator}  \label{supp: nonsep_testbed}

\begin{figure}[ht!]
    \centering
    \includegraphics[width=0.9\textwidth, height=0.85\textheight]{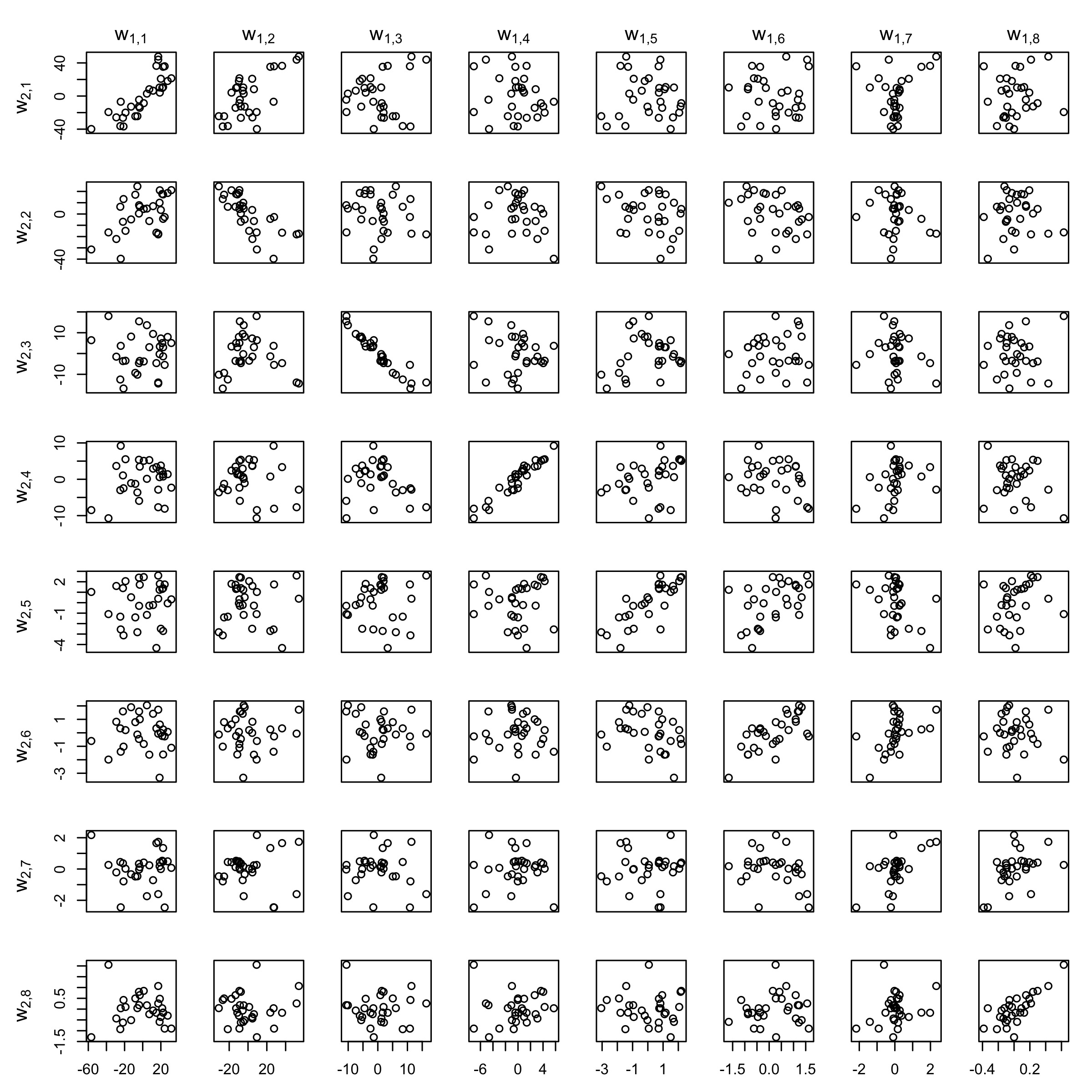}
    \caption{Scatterplot of PC weights for fidelity level $t=1$ (x-axis) and fidelity level $t=2$ (y-axis) evaluated at the inputs in $\mathcal{X}_2$.}
    \label{fig: sim_scatter_wts}
\end{figure}

In the mean function for the $\ell$th level-2 PC weight, we included only the $\ell$th  weight from $t=1$. A scatterplot showing the relationship between each PC weight at the inputs in $\mathcal{X}_2$ for $t=2$ (y-axis) and $t=1$ (x-axis) is provided in Figure \ref{fig: sim_scatter_wts}. The plots along the diagonal of Figure \ref{fig: sim_scatter_wts} show a strong relationship between the $\ell$th  PC weights at successive fidelity levels. We found that including all of the weights from $t=1$ in the mean function for $w_{2,\ell}(\cdot)$ led to a negligible decrease in the RMSPE relative to a model that includes only the $\ell$th  weight from level 1. The scale parameters for the prior assigned to $\theta_{w,t,\ell,i}$ was fixed at half input $x_i's$ range. To stabilize the posterior distribution of $\bftheta_{w,2, 5}$, the scale parameter of $\theta_{w,2,5,i}$ was set to a quarter of variable $x_i's$ range.

\subsection{Visual Comparison for SEP and NONSEP} \label{supp: additional visual comparison}

\begin{figure}[hbt!]
\centering
\includegraphics[width=\linewidth]{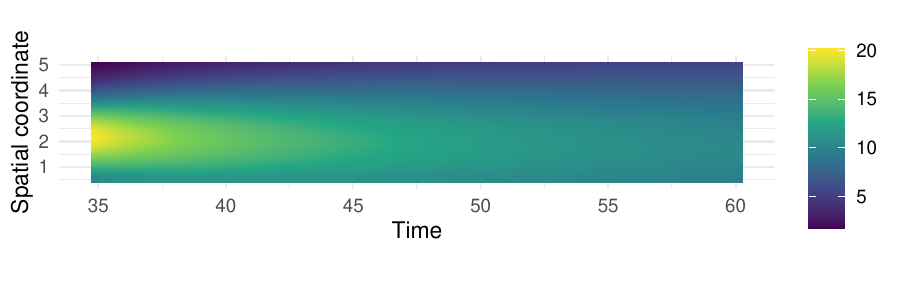}
\vspace{-1.5cm}
\caption{Plot of $y_2(\bfx, \bfs)$ evaluated at the test input $M=11.79$, $D=0.12$, $L=2.18$ and $T=30$ for $\bfs \in [0.5,5] \times [35,60]$.}
\label{fig: sim_function_test}
\end{figure}

\begin{figure}[hbt!]
\centering

\begin{subfigure}[b]{0.49\linewidth}
  \centering
  \includegraphics[width=\linewidth]{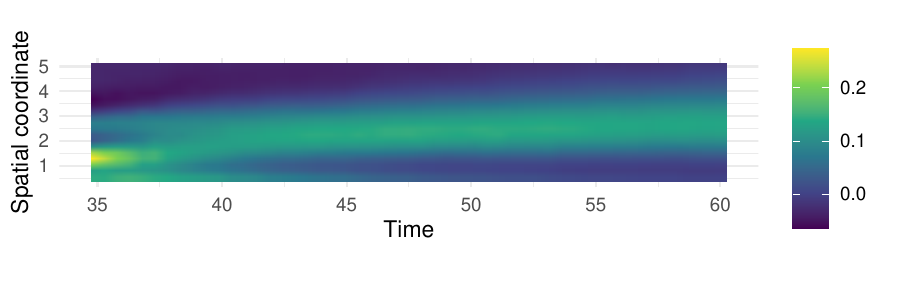}
  \caption{Difference SEP}
  \label{fig: sim_sep_dif}
\end{subfigure}
\hfill
\begin{subfigure}[b]{0.49\linewidth}
  \centering
  \includegraphics[width=\linewidth]{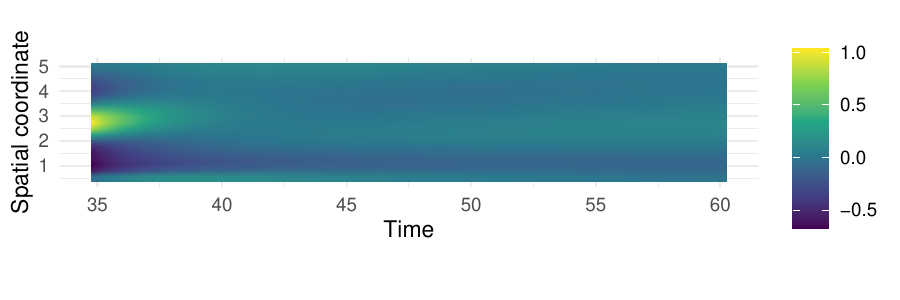}
  \caption{Difference NONSEP}
  \label{fig: sim_nonep_dif}
\end{subfigure}

\begin{subfigure}[b]{0.49\linewidth}
  \centering
  \includegraphics[width=\linewidth]{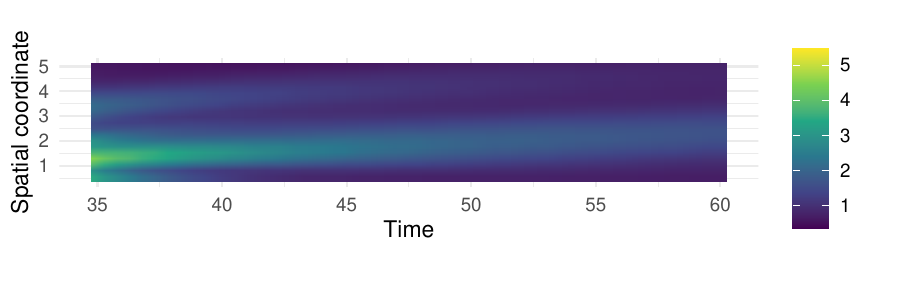}
  \caption{CI Len. SEP}
  \label{fig: sim_sep_ci}
\end{subfigure}
\hfill
\begin{subfigure}[b]{0.49\linewidth}
  \centering
  \includegraphics[width=\linewidth]{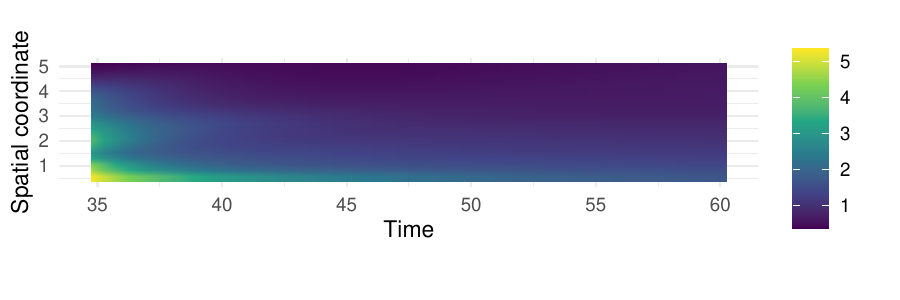}
  \caption{CI Len. NONSEP}
  \label{fig: sim_nonsep_ci}
\end{subfigure}

\caption{Predictive performance of the SEP and NONSEP emulators at the test input $M=11.79$, $D=0.12$, $L=2.18$ and $T=30$. Panels (a) and (b) show the difference between the actual and predicted values of $\bfy_{2}(\bfx, \bfs)$; panels (c) and (d) show $95\%$ credible interval lengths.}
\label{fig: sim_sep_two_level_in1}

\end{figure}

Figure \ref{fig: sim_function_test} shows $\bfy_{2}(\bfx, \bfs)$ over the spatial domain for a single input. Panels (a) and (b) of Figure \ref{fig: sim_sep_two_level_in1} show the difference between the actual and predicted values for the SEP and NONSEP emulators at the test input in Figure \ref{fig: sim_function_test}. Both emulators perform well across the entire space/time domain, but the SEP model has slightly better predictions than the NONSEP model, especially near the region with the greatest output variability. The credible interval lengths are similar for both the SEP and NONSEP emulator; see row 2 of Figure~\ref{fig: sim_sep_two_level_in1}. 

\newpage

\section{Additional Results for the Storm Surge Application} \label{supp: storm surge}

In Section \ref{supp: ss_sep}, we describe several modeling decisions for the SEP emulator, including the number of locations in each neighbor set and the choice of hyperparameters for the inverse cross-covariance matrix at each fidelity level. In Section \ref{supp: ss_nonsep}, we discuss the relationship between the observed PC weights at the two fidelity levels and include several visual diagnostics that are useful for determining the number of principal components to include in the model.

\subsection{Diagnostics and Implementation Details for the SEP Emulator} \label{supp: ss_sep}

\begin{figure}[hbt!]
    \centering
    \begin{subfigure}[b]{0.32\textwidth}
        \centering
        \includegraphics[width=\textwidth]{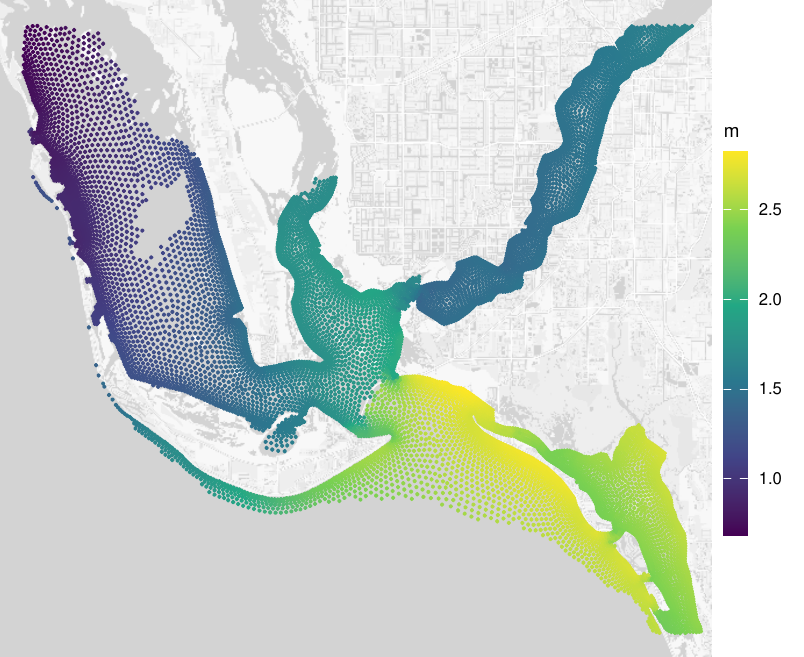}
        \caption{Mean ADCIRC}
        \label{fig: ss_mean_L1}
    \end{subfigure}
    \hfill
    \begin{subfigure}[b]{0.32\textwidth}
        \centering
        \includegraphics[width=\textwidth]{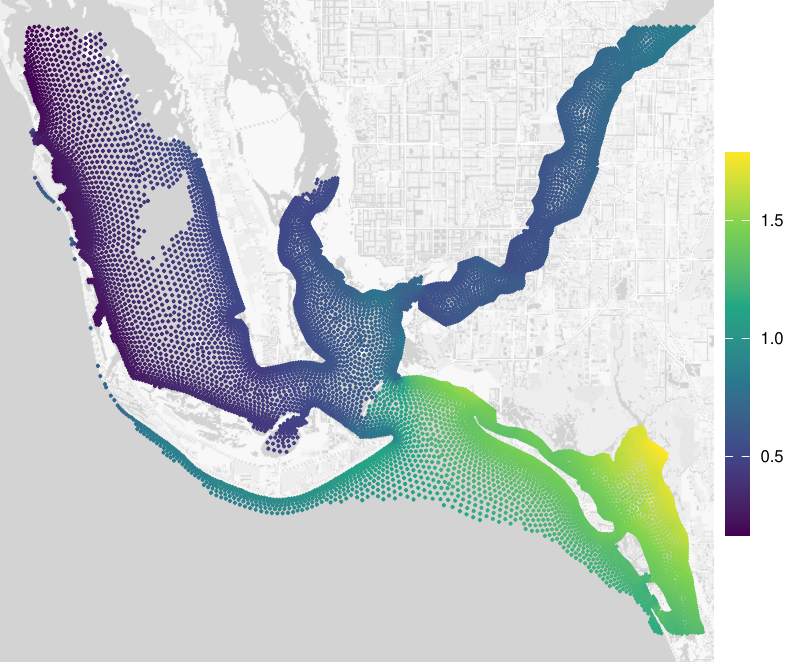}
        \caption{Variance ADCIRC}
        \label{fig: ss_var_L1}
    \end{subfigure}
    \hfill
    \begin{subfigure}[b]{0.32\textwidth}
        \centering
        \includegraphics[width=\textwidth]{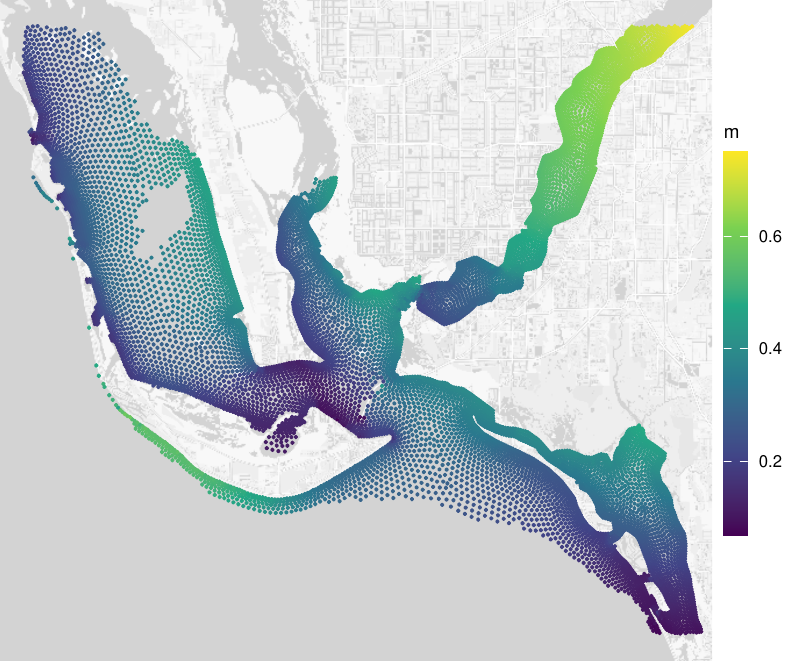}
        \caption{Residual ADCIRC}
        \label{fig: ss_res_L1}
    \end{subfigure}

    \vspace{0.5cm}

    \begin{subfigure}[b]{0.32\textwidth}
        \centering
        \includegraphics[width=\textwidth]{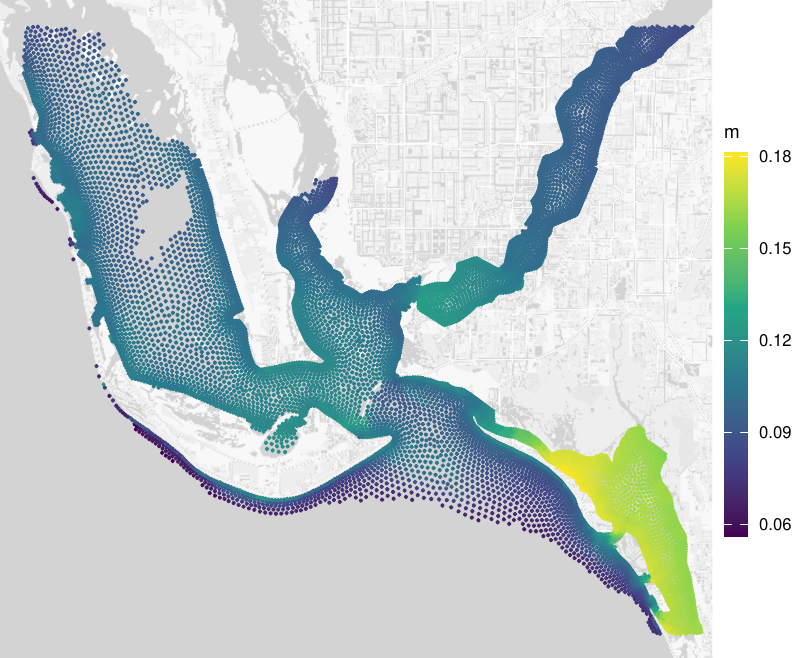}
        \caption{Mean ADCIRC + SWAN}
        \label{fig: ss_mean_L2}
    \end{subfigure}
    \hfill
    \begin{subfigure}[b]{0.32\textwidth}
        \centering
        \includegraphics[width=\textwidth]{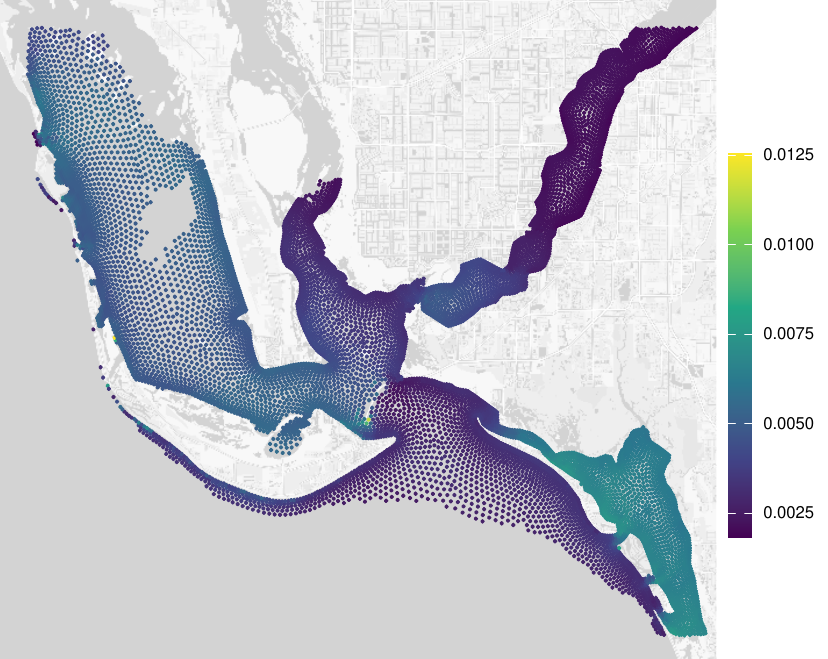}
        \caption{Variance ADCIRC + SWAN}
        \label{fig: ss_var_L2}
    \end{subfigure}
    \hfill
    \begin{subfigure}[b]{0.32\textwidth}
        \centering
        \includegraphics[width=\textwidth]{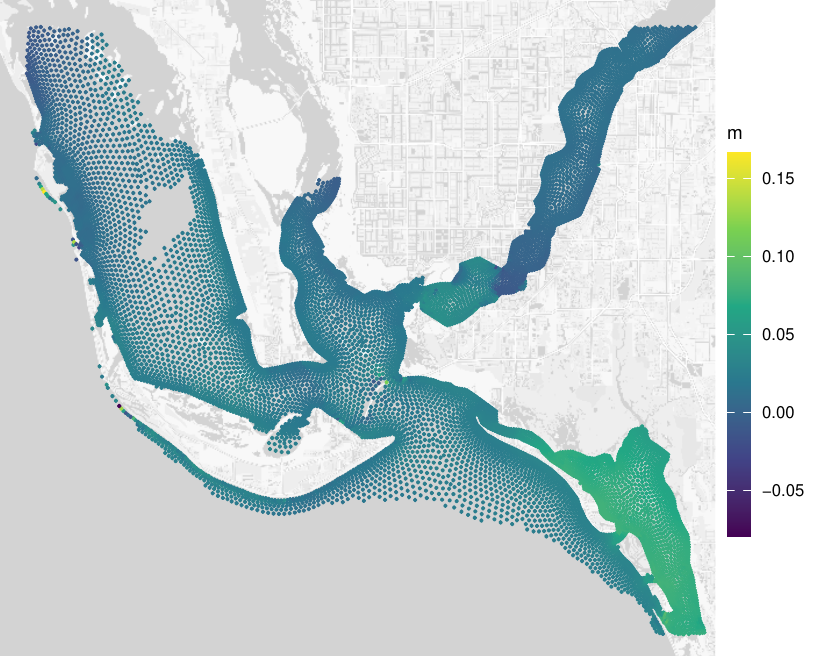}
        \caption{Residual ADCIRC + SWAN}
        \label{fig: ss_res_L2}
    \end{subfigure}
    
    \caption{Comparison of the spatial trend and variability for the ADCIRC and ADCRIC + SWAN outputs. Panels (a) and (b) show the empirical mean and variance of ADCIRC outputs evaluated at the inputs in $\mathcal{X}_1$. Panels (d) and (e) show the empirical mean and variance for the difference between the ADCIRC + SWAN and ADCIRC outputs over $\mathcal{X}_2$. Panels (c) and (f) show the residuals for the two fidelity levels after removing the spatial trend at the input $\Delta P=49.82$, $R_p=19.47$, $V_f=5.86$, $\theta=50.23$, $B =1.02$ and $\ell=(-82.20, 26.75)$.}
    \label{fig: ss_mean_var_res}
\end{figure}

Panels (a), (b), (d) and (e) of Figure \ref{fig: ss_mean_var_res} show the empirical mean and variance of the ADCIRC and ADCIRC + SWAN outputs, indicating nonstationarity at both fidelity levels. Based on the difference in variance between panels (b) and (e), the level-1 outputs explain much of the variability in the level-2 outputs, suggesting that an autoregressive structure is suitable for a model of the ADCIRC + SWAN outputs. After removing the spatial trend from the outputs, Panels (c) and (f) indicate that the residuals at each fidelity level still exhibit a high degree of spatial correlation.

\begin{figure}
    \centering
    \begin{subfigure}[b]{0.49\linewidth}
        \includegraphics[width=\linewidth]{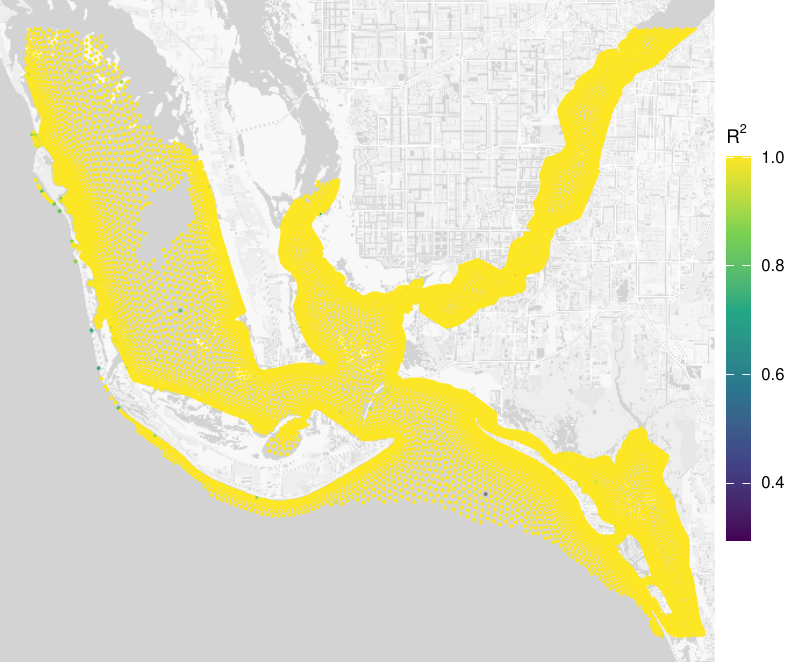}
    \caption{ADICRC}
    \label{fig: ss_r_sq_L1}
    \end{subfigure}
    \hfill
    \begin{subfigure}[b]{0.49\linewidth}
        \includegraphics[width=\linewidth]{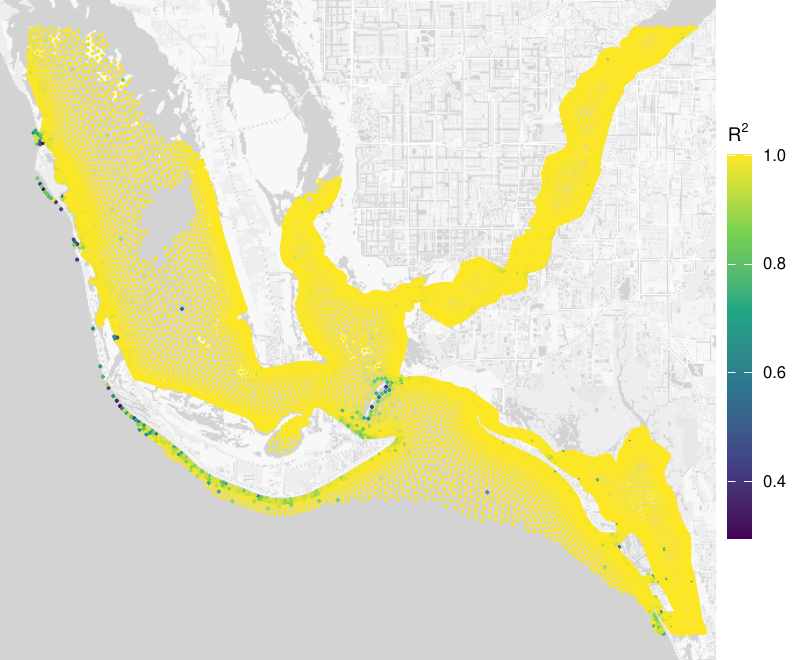}
        \caption{ADCIRC + SWAN}
        \label{fig: ss_r_sq_L2}
    \end{subfigure}
    \caption{Output variability by location explained by the outputs of a nearest neighbor. Panel (a) shows the the coefficient of determination ($R^{2}$) for the centered outputs at a location and its nearest neighbor for the ADCIRC simulator. Panel (b) shows the coefficient of determination for the centered ADCIRC + SWAN outputs after subtracting the ADCIRC outputs.}
    \label{fig: ss_r_sq}
\end{figure}

Based on the plots in Panels (a) and (c), the basis function $h_{t}(\bfx)$ was fixed at $1$ for $t=1,2$. After sorting the spatial locations using the maximum-minimum distance ordering \citep{Guinness_sharpening}, the neighbor set $\Cjset$ for $t=1$ and $t=2$ was constructed using the index of the nearest point to $\bfs_j$ from $\bfs_{1}, \ldots, \bfs_{j-1}$. Figure \ref{fig: ss_r_sq} shows the percentage of explained variability from regressing the centered outputs at a location on the centered outputs of the location's nearest neighbor. A single neighbor explains nearly all the residual variability at most locations. Because neighboring outputs are highly correlated, when $\Cjset$ includes more than one neighbor, the centered columns of $\bfy_{t, \Cjset}$ are nearly linearly dependent for many locations, causing numerical problems when evaluating $p(\bftheta |\yObs)$.

Conditional on $d_{t,j}$, $\bfa_{t,j}^{(\Cjset)}$ was assigned a weakly informative prior with the hyperparameter $\tau^{2}_t = 10^{7}$ for $t=1,2$. The conditional variance parameter $d_{t,j}$ was also given a vague prior, with $\eta_{t}=0.2$ for $t=1,2$, $\lambda_1=0.02$ and $\lambda_2 = 0.002$. To obtain samples from the posterior distribution, the Metropolis-Hastings algorithm was run for $30,000$ iterations (with a burn-in of $3,000$ samples), with every tenth sample saved for predictions.

Because neighboring outputs are highly correlated at both fidelity levels, the distribution of $p(\bftheta | \yObs)$ is particularly sensitive to the choice of $\lambda_t$. Small $\lambda_t$ penalizes large values of $\bftheta_t$ in the posterior distribution $p(\bftheta_t |\yObs)$. If $\lambda_t$ is too small, most of the mass for $p(\bftheta_t | \yObs)$ concentrates in a region close to zero, leading to poor predictive performance for the emulator. At $t=2$, for example, when $\tau^{2}_2 = 10^7$, $\eta_2 = 0.2$ and $\lambda_2 = 0$ (these hyperparameters correspond to the improper prior $p(d_{2,j}) \propto d_{2,j}^{-1.1})$, $p(\bftheta_2 | \yObs)$ is maximized at $\bftheta_2 = \bfzero$. As $\lambda_t$ increases, the penalty for large $\bftheta_t$ fades, leading to a ridge in the posterior distribution of $\bftheta_t$, making it difficult to implement a sampler or perform MAP estimation. To penalize large values of $\bftheta_t$, the hyperparameter $\lambda_t$ must be fixed close to zero. It could be interesting to investigate if the assignment of the jointly robust prior \citep{Gu2019} to the range parameters improves MAP estimation of $\bftheta$ for a wider range of $\lambda_t$. 

To assess the sensitivity of the SEP model to $\lambda_1$, we computed predictions using the MAP estimator of $\bftheta$ for several values of $\lambda_1$ (see Table \ref{tab: ss_hp_selection}). The remaining hyperparameters in the priors for the cross-covariance matrices were fixed at $\tau^{2}_t=10^7, \eta_t =0.2$ for $t=1,2$ and $\lambda_2=0.002$. The results in Table \ref{tab: ss_hp_selection} are similar for other small values of $\lambda_2$ because the ADCIRC simulations explain a high percentage of the ADCIRC + SWAN output variability. As $\lambda_1$ increases, the MAP estimate of $\bftheta_1$ increases for each input variable. The predictive variance in Table \ref{tab: ss_hp_selection} decreases as a function of $\lambda_1$ because larger values of $\bftheta_1$ shrink the term $R_{1}(\bfx_0)$, defined in \eqref{def: R_t_star_star}, which quantifies predictive uncertainty for a new input at all locations.

\begin{table}[hbt!]
\centering
\caption{Predictive performance of the SEP emulator as a function of $\lambda_1$. The SEP model was estimated using the MAP estimator of $\bftheta$ and assessed on $106$ test inputs.}
\label{tab: ss_hp_selection}
\begin{tabular}{cccccc}
\toprule
  & \multicolumn{3}{c}{Marginal predictive performance of $y_{2,j}(\cdot)$} & \multicolumn{2}{c}{Predictive performance of $\Bar{y}_2(\cdot)$} \\
\cmidrule(lr){2-4} \cmidrule(lr){5-6}
$\lambda_1$ & RMSPE & CVG ($95\%$) & ALCI ($95\%$) & CVG $(95\%$) &  ALCI ($95\%$) \\ 
  \hline
0.002 & 0.137 & 97.89 & 0.613 & 96.23 & 0.509 \\  
  0.02 & 0.111 & 90.61 & 0.394 & 89.62 & 0.275 \\ 
  0.1 & 0.114 & 84.59 & 0.347 & 82.08 & 0.214 \\ 
   \hline
\end{tabular}
\end{table}

\subsection{Diagnostics and Implementation Details for the NONSEP Emulator} \label{supp: ss_nonsep}

\begin{figure}[ht!]
    \centering
    \includegraphics[width=\textwidth, height=0.9\textheight]{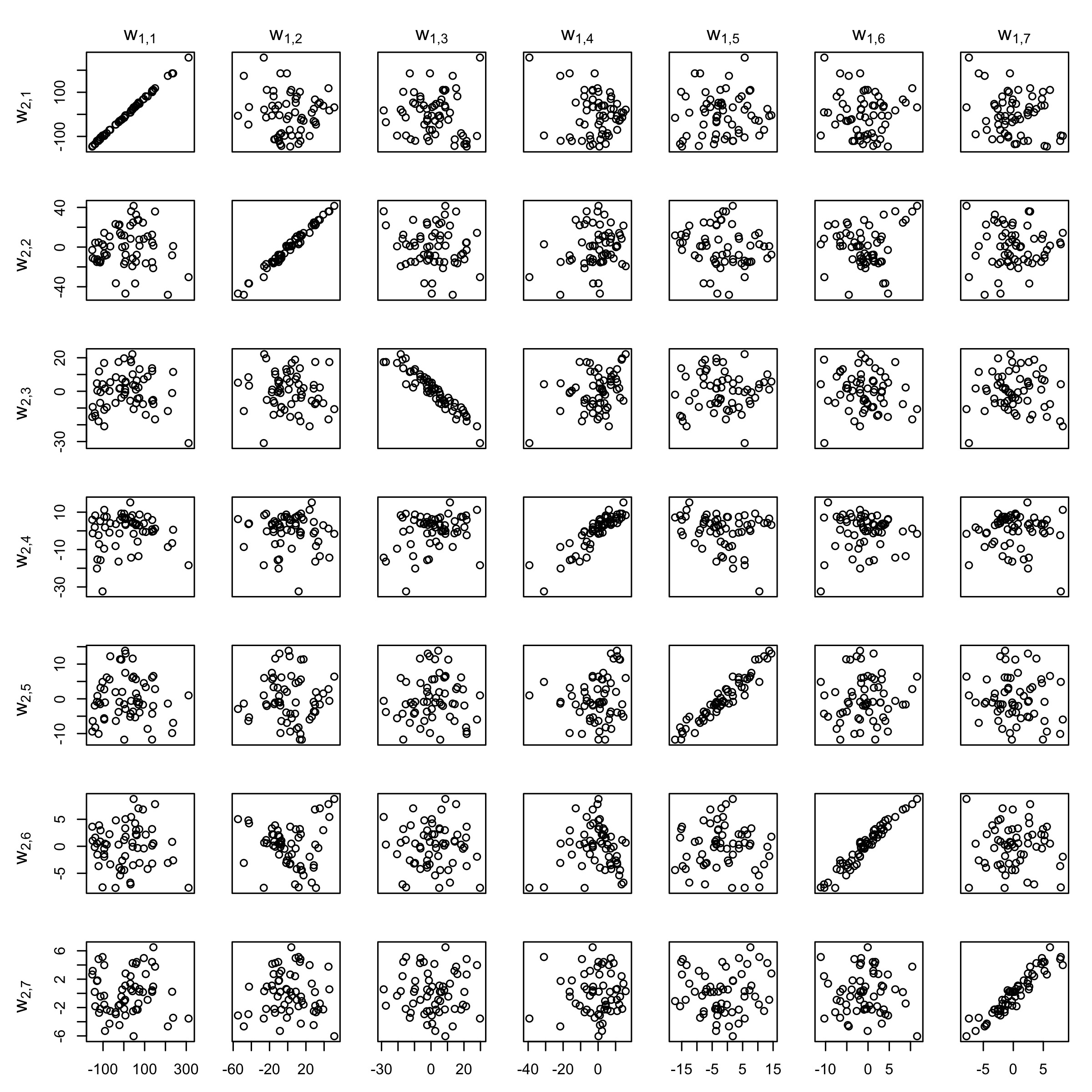}
    \caption{Scatterplot of PC weights at fidelity level $t=1$ (x-axis) and fidelity level $t=2$ (y-axis) evaluated at the inputs in $\mathcal{X}_2$.}
    \label{fig: storm surge PC weights}
\end{figure}

Figure \ref{fig: storm surge PC weights} shows the relationship between the PC weights for the ADCIRC and ADCIRC + SWAN simulators. Because the ADCIRC simulation is a good approximation to the ADCIRC + SWAN model, the $\ell^{\text{th}}$ weights at the two fidelity levels are highly correlated. Along the diagonal of Figure \ref{fig: storm surge PC weights}, the level-2 and level-1 weights are positively correlated, except for PC three. The direction of correlation for the weights at the third principal component is negative because the basis vectors in \eqref{def: basis_rep} are unique up to a sign. Based on the relationship between the PC weights over the two fidelity levels shown in Figure \ref{fig: storm surge PC weights}, the mean function for $\bfw_{2,\ell}(\cdot)$ was simplified to only include the $\ell^{\text{th}}$ weight from level 1. After specifying the model for the random weights, the scale parameters for the prior assigned to $\theta_{w,t,\ell,i}$ were fixed at half variable $x_i$'s range. For each PC weight and fidelity level, the Metropolis-Hastings algorithm was run for $60,000$ iterations (with a burn-in of $6,000$ samples), with every tenth sample saved for predictions.

\begin{figure}[hbt!]
    \centering
    \begin{subfigure}[b]{0.49\textwidth}
        \centering
        \includegraphics[width=\textwidth]{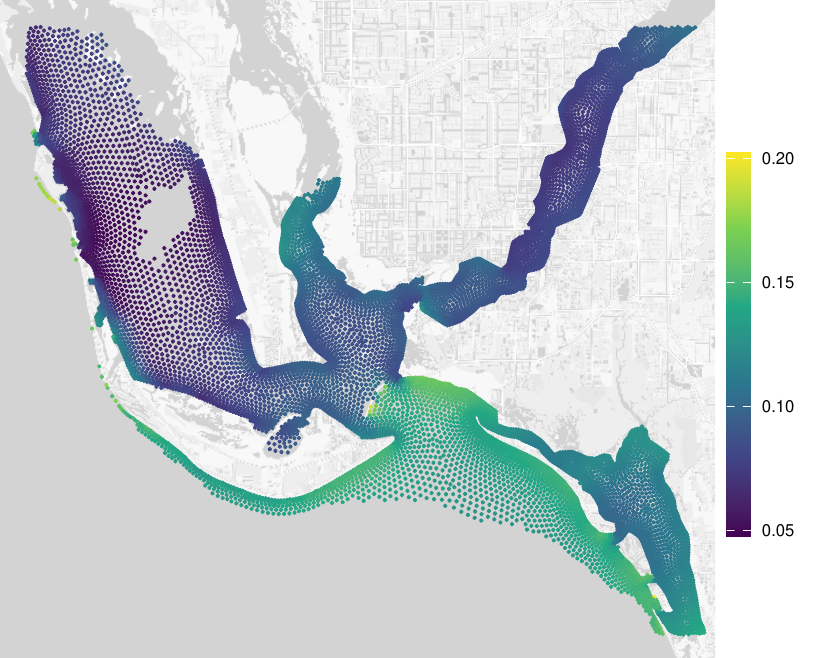}
        \caption{RMSPE 6 PCs.}
    \end{subfigure}
    \hfill
    \begin{subfigure}[b]{0.49\textwidth}
        \centering
        \includegraphics[width=\textwidth]{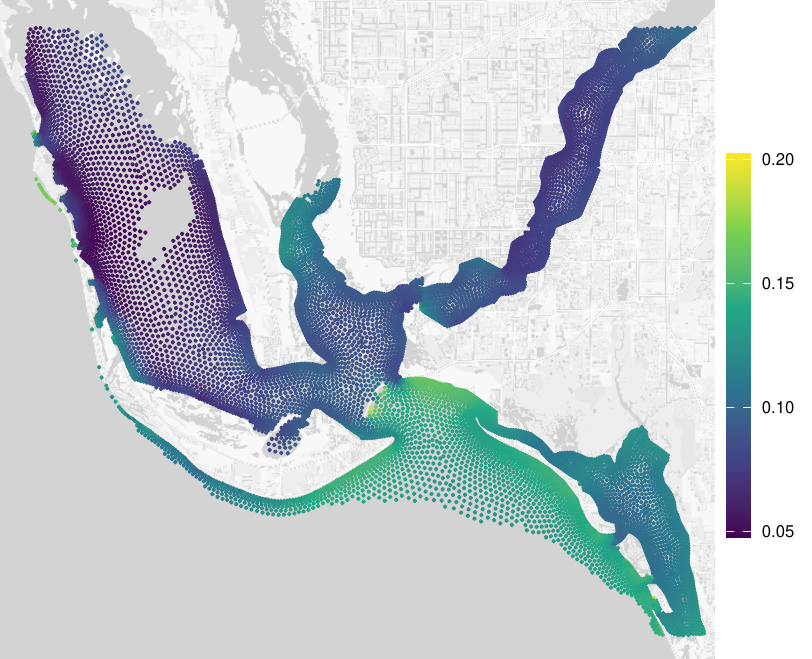}
        \caption{RMSPE 7 PCs.}
    \end{subfigure}
    \vspace{0.5cm}
    \begin{subfigure}[b]{0.49\textwidth}
        \centering
        \includegraphics[width=\textwidth]{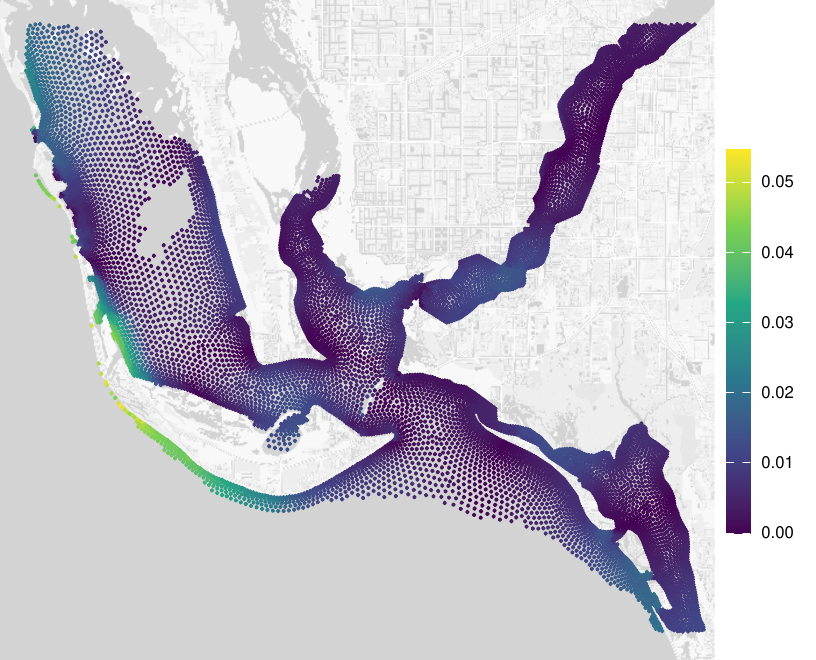}
        \caption{$|\bfk_{2,7}|$}
    \end{subfigure}
    
    \caption{Maps of RMSPE for models with different numbers of principal components. Panels (a) and (b) show the RMSPE over testing inputs when the NONSEP model includes six and seven PCs, respectively. Panel (c) shows the absolute values of the basis vector $\bfk_{2,7}$ in Equation~\eqref{def: basis_rep}. }
    \label{fig: ss_rmspe_pc_6-7}
\end{figure}

Figure \ref{fig: ss_rmspe_pc_6-7} shows the RMSPE by location for models that include six or seven principal components. Panels (a) and (b) show that the seventh principal component hardly changes the RMSPE at most locations. But in a small region along the southwestern coast, where the absolute values of $\bfk_{2,7}$ are greatest, the addition of the seventh PC weight leads to a meaningful reduction in the RMSPE. 

\end{document}